\documentclass{article}
\pdfoutput=1
\usepackage{macros}
\preprint{
{\tt DESY-23-215}
}
\title{Unravelling T-Duality: Magnetic Quivers in Rank-zero Little String Theories
}
\author[a]{Lorenzo Mansi}
\author[b]{and Marcus Sperling}
\affiliation[a]{Deutsches Elektronen-Synchrotron DESY, \\ Notkestr. 85, 22607 Hamburg, Germany}
\affiliation[b]{Fakultät für Physik, Universität Wien,\\
Boltzmanngasse 5, 1090 Wien, Austria}
\emailAdd{lorenzo.mansi@desy.de}
\emailAdd{marcus.sperling@univie.ac.at}

\abstract{
An intriguing class of 6d supersymmetric theories are known as little strings theories, which exhibit a rich network of T-dualities. A robust feature of these theories are their Higgs branches. Focusing on the little string theories that are realised on a single curve of zero self-intersection, we utilise brane systems to derive the magnetic quivers. Using a variety of techniques (including branching rules, brane dynamics, F-theory geometry, quiver subtraction, and the decay and fission algorithm), we detail the Higgs branch Hasse diagram and determine the transverse slices for every elementary Higgs branch RG-flow. Building on these insights, we pursue two directions: firstly, we used the established connection between the change of the 2-Group structure constants along Higgs branch RG-flows and the transition-type in the Hasse diagram to infer putative T-dual models. Secondly, we conjecture an algorithm that predicts the non-Abelian flavour symmetry of the compactified little string theory by inspecting the magnetic quivers of all T-dual frames.  
}

\begin{document}

\maketitle

\section{Introduction}\label{sec:Introduction}
The landscape of six-dimensional supersymmetric theories has been explored from different perspectives; including field theory 
\cite{Ohmori:2014kda, Ohmori:2014pca,Cordova:2015fha,Shimizu:2017kzs,Chang:2017xmr},
geometry \cite{Aspinwall:1996nk,Aspinwall:1997eh,Aspinwall:1996vc,Morrison:1996na,Morrison:1996pp,Aspinwall:1997ye,Morrison:2012np,Heckman:2015bfa},
 and brane constructions \cite{Hanany:1997gh,Intriligator:1997kq,Intriligator:1997dh,Brunner:1997gf},
 and subsequent works. There are three possible kinds of theories that can be engineered in six dimensions: super-gravity theories (SUGRA), superconformal field theories (SCFTs), and little string theories (LSTs); the latter have recently re-emerged as a prominent subject of investigation \cite{Witten:1995zh, Aspinwall:1996vc, Aspinwall:1997ye, Intriligator:1997dh, Seiberg:1997zk, Bhardwaj:2015oru, DelZotto:2020sop, DelZotto:2022ohj, DelZotto:2022xrh, DelZotto:2023ahf, DelZotto:2023nrb,Ahmed:2023lhj,Lawrie:2023uiu}.

Conventional approaches might be insufficient to study LSTs due to their non-gravitational and non-local nature. A quantum field theory (QFT) description is only viable at energies $E \le T_{\mathrm{LST}} $, where $T_{\mathrm{LST}}$ represents the inherited little string tension.
Moreover, little string theories exhibit T-duality which is notoriously difficult to track: the choice of Wilson lines during the circle compactification is a piece of data that is not explicitly manifest in the theory.
Nevertheless, there has been a resurgence of interest towards these theories since it has been realised that the advancements in the study of higher form symmetries, stemming from the original paper \cite{Gaiotto:2014kfa}, allowed to shed light on T-duality properties of LSTs which had only been probed in earlier geometric approaches  \cite{Morrison:1996na,Morrison:1996pp}.

The pioneering program initiated in \cite{DelZotto:2020sop, DelZotto:2022ohj, DelZotto:2022xrh, DelZotto:2023ahf} relies on matching T-duality invariant quantities to identify potential pairs of dual theories. Along with the classical invariants, given by the rank of the flavour symmetry group, and the Coulomb branch dimension, physicists know very well that anomaly matching plays a key role in identifying dual theories. In fact, restricting to the class of $\mathcal{N}=\left( 1,0 \right)$ LSTs, these models always manifest a 2-Group global symmetry\footnote{The challenge lies in the central $\urm(1)$ one-form extension to a zero-form symmetry. This does not always ensure that the background gauge transformation, mixing the one-form symmetry with the zero-form symmetries, aligns with the transformation law of a 2-connection on a principal smooth 2-Group bundle \cite{Kang:2023uvm}. 
But since we are only interested in the mixing between the one-form symmetry and the zero-form symmetry, we continue to refer to \eqref{eqn:2-Group} as the 2-Group symmetry of the theory, without worrying about whether the associated background gauge transformation gives a faithful transformation law or not.}
\begin{equation}\label{eqn:2-Group}
    2\text{-}\mathrm{Grp}=\left( \mathscr{P}^{(0)} \times \sprm(1)^{(0)}_R \times F^{(0)} \right) \times_{\kappa_{\mathscr{P}} , \kappa_R , \kappa_{F}} \urm(1)^{(1)}_{LST} \ , 
\end{equation}
that originates from a mixing of the 0-forms Poincar\'e $\mathscr{P}^{(0)}$, R-symmetry $\sprm(1)^{(0)}_R$, and flavour symmetry $F^{(0)}$ with the 1-form $ \urm(1)^{(1)}_{LST}$, associated with the LST charge. The 2-Group structure constants $\kappa_{\mathscr{P}}$, $\kappa_R$, $\kappa_F$ regulate the interplay between each 0-form symmetry and the 1-form symmetry under background gauge transformation. Thus, they can be thought of as anomaly coefficients coming from a box diagram with two external dynamical gauge gluons, two external 1-form currents associated with the background 0-form symmetries, and fermions running in the loop \cite{Cordova:2020tij}.

Together with the data coming from these higher-form symmetry anomalies, the full list of invariant quantities under T-duality is:
\begin{equation} \label{eqn:T-duality_Invariants}
    \mathrm{dim}(\mathcal{C}) \quad , \quad \mathrm{rank}(\mathfrak{f}) \quad , \quad \kappa_{\mathscr{P}} \quad , \quad  \kappa_R  \quad , \quad \kappa_{F} \,.
\end{equation}
Here, $\mathrm{dim}(\mathcal{C})$ is the dimension of the Coulomb branch of the $5d$ supersymmetric theory obtained via $S^1$ reduction, whereas the rank of the six-dimensional flavour symmetry $\mathrm{rank}(\mathfrak{f})$ corresponds to the dimension of the space of Wilson Line that can be turned in the compactification procedure. One therefore requires candidate T-dual pairs to have the same set of invariants \eqref{eqn:T-duality_Invariants} \cite{DelZotto:2022ohj,DelZotto:2020sop,Ahmed:2023lhj}. 

To avoid undergoing the computation of each of the quantities in \eqref{eqn:T-duality_Invariants} for every theory one can engineer, it is possible to recur to renormalisation group (RG) flows on the Higgs branch of the LSTs to automatically extract putative T-dual pairs having the same invariants \cite{Lawrie:2023uiu}. 

Since the (Higgs branch) Hasse diagram structure automatically enforces the matching of the aforementioned quantities in a systematic way, we are thus interested in the study of the Higgs branches of little string theories. Considering models with minimal supersymmetry  --- i.e.\ 8 supercharges in six dimensions --- we can deploy the \emph{magnetic quiver} technique to extract the sought-after moduli space 
\cite{Cabrera:2019izd,Cabrera:2019dob,Sperling:2021fcf,Hanany:2022itc,Fazzi:2022hal,Bourget:2022tmw,Fazzi:2022yca}.
That is, a finite collection $\mathrm{Q}_i$ of $3d$ $\mathcal{N}=4$ (generalised) quiver theories such that the union of their Coulomb branches $\mathcal{C}(\mathrm{Q}_i)$ provides a geometric description of the Higgs branch $\mathcal{H}$ of the original six-dimensional theory is engineered:
\begin{equation}
    \mathcal{H}_{6d} \cong \bigcup \limits_{i} \bigg( \mathcal{C}_{3d} (\mathrm{Q}_i) \bigg) \,.
\end{equation}
The power of this construction is that it does not require performing a $T^3$ compactification and then employing 3d mirror symmetry, but manages to extract such quiver theory directly from the six-dimensional model when the latter has a realisation in Type IIA superstring theory \cite{Cabrera:2019izd,Cabrera:2019dob,Sperling:2021fcf,Hanany:2022itc,Fazzi:2022hal,Bourget:2022tmw,Fazzi:2022yca}.

Along with the magnetic quiver techniques comes the Higgs branch Hasse diagram \cite{Bourget:2019aer}, which encodes the finite stratification of $\mathcal{H}_{6d}$ into symplectic leaves.  Given a (unitary) magnetic quiver for $\mathcal{H}_{6d}$, the Hasse diagram can be derived via two independent algorithms: (i) \emph{quiver subtraction} \cite{Cabrera:2018ann,Bourget:2019aer,Bourget:2020mez,Bourget:2022ehw,Bourget:2022tmw} and (ii) \emph{decay and fission} \cite{Bourget:2023dkj,InvQuiverSubHiggs}. The former relies on a list of minimal transitions (which is assumed to be complete) and generates magnetic quivers for the (closures) of the symplectic leaves, but not for the theories after Higgsing. The decay and fission algorithm, on the other hand, does not rely on any prior input and generates the entire Hasse diagram holistically from the magnetic quiver. In particular, the quivers generated are the magnetic quivers for the theories living at the end of the Higgs branch RG-flow. As a consequence, combining both algorithms provides unprecedented insights into the Higgs branch geometry and the RG-flows.

In this paper, we focus on a little string theory with a sole non-dynamical tensor paired with a gauge algebra of type special unitary or symplectic, for which a D6-NS5-D8 brane realisation in Type IIA superstring theory exists. 
In particular, we extract the magnetic quivers for this specific family of models and derive the associated (Higgs branch) Hasse diagrams. Later on, we investigate possible T-duals for this model and propose a rule to read the maximal non-Abelian symmetry we could expect upon compactification directly from the six-dimensional theory.  

\paragraph{Outline.}
The remainder is organised as follows:  after a brief review of little string theories in Section~\ref{sec:Brief_LST}, Section~\ref{sec:Branes-MQ} focuses on the brane realisation of these theories and the derivation of the magnetic quivers. Building on this, Higgs branch RG-flows are deduced via a variety of techniques, including brane motions, branching rules in the low energy effective description, quiver subtraction, and the decay and fission algorithm. In Section~\ref{sec:F-theory}, the perspective is changed and the little string theories are analysed based on their F-theory realisation. Lastly, conclusions are provided in Section~\ref{sec:conclusions}. 
The main body is complemented by Appendix~\ref{app:monopole}, which details technical aspects.

\paragraph{Notation.}
A unitary \emph{magnetic quiver} is a graph composed of vertices and edges that encodes a (generalised) $3d$ $\mathcal{N}=4$ quiver theory by specifying:
\begin{compactitem}
    \item A vertex $\raisebox{-.5\height}{
 \includegraphics[page=15]{pics/0-curve_IQS_figures.pdf}
 }$ denotes a $\urm(N)$ vector multiplet. 
    \item An edge $\raisebox{-.5\height}{
 \includegraphics[page=16]{pics/0-curve_IQS_figures.pdf}
 }$ denotes a hypermultiplets with multiplicity $m$ and non-simply lacedness $\ell$. For $\ell =1$, this reduces to $m$ hypermultiplets in the bifundamental representation of $\urm(N_1) \times \urm(N_2)$. For $m=1$ and $\ell >1$, this reduces to the non-simply laced edge introduced in \cite{Cremonesi:2014xha}.
\end{compactitem}
In some cases, quivers need to be ``decorated'' \cite{Bourget:2022ehw} to account for repeated identical transitions in quiver subtraction, see below. 
A vertex (unitary gauge node) is called \emph{balanced} if the number of flavours equals twice the rank.

A \emph{Hasse diagram} is a graph that encodes a finite partially ordered set. For symplectic singularities $X$ \cite{beauville2000symplectic}, such as Higgs branches of theories with 8 supercharges or the 3d $\mathcal{N}=4$ Coulomb branch, a Hasse diagram is used to encoded the finite canonical stratification \cite{kaledin2006symplectic}: $X =\overline{\mathcal{L}}_1\supset \overline{\mathcal{L}}_2 \supset \cdots$. The basic concepts are
\begin{compactitem}
    \item Vertices denote \ul{symplectic leaves} $\mathcal{L}_i$. The trivial leaf $\mathcal{L}_{\emptyset}$ is at the bottom, while the highest dimensional leaf $\mathcal{L}_1$ is at the top.
    \item Edges denote \ul{minimal transitions/degenerations}; i.e.\ two ordered leaves $\mathcal{L}_{i,j}$ with $\mathcal{L}_{i}< \mathcal{L}_{j}$ are a minimal degeneration if there does not exists another leaf $\mathcal{L}_k \neq \mathcal{L}_{i,j}$ with $\mathcal{L}_{i}< \mathcal{L}_{k}< \mathcal{L}_{j}$.
\end{compactitem} 
Between any two ordered leaves $\mathcal{L}_{i,j}$, with $\mathcal{L}_{i}< \mathcal{L}_{j}$, there exists a transverse slice $\mathcal{S}_{i,j}$, which is the transverse space to a point in $\mathcal{L}_{i}$ inside the closure $\overline{\mathcal{L}}_{j}$.
Connecting with magnetic quivers, there are two distinct algorithms that deduce the Hasse diagram:
\begin{compactitem}
    \item \ul{Quiver subtraction} \cite{Cabrera:2018ann,Bourget:2019aer,Bourget:2020mez,Bourget:2022ehw,Bourget:2022tmw}: Derives magnetic quivers for the transverse slices $S_{\emptyset,i} \cong \overline{\mathcal{L}}_i$ from the bottom.

    \item \ul{Decay and fission} \cite{Bourget:2023dkj,InvQuiverSubHiggs}: Derives magnetic quivers for the transverse slices $S_{i,1}$ to the top.
\end{compactitem}

\section{A brief introduction to LSTs} 
\label{sec:Brief_LST}
We briefly review some characteristics of little string theories. 
The general toolkit used to engineer six-dimensional theories consists of F-theory upon a suitable compactification on an elliptically fibred Calabi-Yau threefold $\mathrm{CY}_3$, the specifics of this space determine the resulting lower dimensional theory. As such, LSTs can be constructed by choosing a non-compact base space $\mathcal{B}$ wherein complex curves $C_i$ live; these curves intersect pairwise and can be decorated with singular fibres fitting the Koidara classification. How the curves intersect themselves is encoded in the intersection matrix $\eta=C_i \cdot C_j$, which is square symmetric by construction. The key point for the realisation of an LST is that there \emph{must} exist a cycle of finite size in the base space that cannot be shrunk to zero size. This translates into the requirement that $\eta$ has \emph{one and only one} null eigenvalue and all its minors are negative defined. 

It is therefore possible to describe an LST, much like to what has been done for SCFTs \cite{Heckman:2015bfa,Heckman:2013pva}, just by specifying the curve configuration along with a few minimal constituents. The common notation used in the literature is:
\begin{equation}
    \stackunder{\stackon{$n$}{$\mathfrak{g}$}}{$ [N_\rho]$} 
\end{equation}
where $n$ is minus the self-intersection number of the associated compact curve $C$, $\mathfrak{g}$ is the gauge algebra dictated by the fibration\footnote{The correspondence between gauge algebras and singular fibres is, in general, not bijective. But, in the models considered in this paper the fibres are always of type $I_n$, such that there is no ambiguity in the association.}, and $N_\rho$ is the number of hypermultiplets in the representation $\rho$ of $\mathfrak{g}$ necessary to cancel gauge anomalies\footnote{See \cite[Tab.\ 3]{Heckman:2018jxk} for a full list of gauge algebras supported on a $-n$ curve together with the enforced matter content.}. 

More complex theories can be constructed from the following \emph{minimal} theories:
\begin{equation}\label{eqn:building_blocks}
   \begin{gathered}
      \overset{\mathfrak{su}_3}{3} \quad , \quad \overset{\mathfrak{so}_8}{4} \quad , \quad \overset{\mathfrak{f}_4}{5} \quad , \quad \overset{\mathfrak{e}_6}{6} \quad , \quad \overset{\mathfrak{e}_7}{7} \,, \quad \overset{\mathfrak{e}_7}{8} \quad , \quad \overset{\mathfrak{e}_{8}}{12} \,, \quad \overset{\mathfrak{su}_{2}}{2}\overset{\mathfrak{g}_{2}}{3} \quad , \quad 2\overset{\mathfrak{su}_{2}}{2}\overset{\mathfrak{g}_{2}}{3} \quad , \quad \overset{\mathfrak{su}_{2}}{2}\overset{\mathfrak{so}_{7}}{3}\overset{\mathfrak{su}_{2}}{2} \ , \\
    \underbrace{2\cdots 2}_{N-1} \quad , \quad 
    \underbrace{2\cdots 2}_{N-3}\overset{\displaystyle 2}{2}2 \quad , \quad 
    22\overset{\displaystyle 2}{2}22 \quad , \quad 
    222\overset{\displaystyle 2}{2}22 \quad , \quad 
    2222\overset{\displaystyle 2}{2}22 \ ,
    \end{gathered}
\end{equation}
where the first line in \eqref{eqn:building_blocks} comprises the minimally decorated $6d$ $\mathcal{N}=(1,0)$ SCFTs, also known as non-Higgables clusters (NHCs)\cite{Morrison:2012np}, having trivial Higgs branch, whereas the second line contains the $6d$ $\mathcal{N}=(2,0)$ ADE clusters of undecorated $-2$ curves. Then non-minimal theories are obtained by \emph{gluing} two minimal theories together with  an E-string theory, in our language a $-1$ curve, in the middle:
\begin{equation}
     \stackon{$m$}{$\mathfrak{g}_L$} \text{ and } \stackon{$n$}{$\mathfrak{g}_R$}  \xrightarrow[]{\text{gluing}} \stackon{$m$}{$\mathfrak{g}_L$} \stackunder{$1$}{$[\mathfrak{f}_e]$}\stackon{$n$}{$\mathfrak{g}_R$} \qquad \mathrm{iff} \quad \mathfrak{g}_L \oplus \mathfrak{g}_R \subseteq E_8 \ , \ \mathfrak{f}_e= \mathrm{Comm}\left(\mathfrak{g}_L \oplus \mathfrak{g}_R , E_8\right) \ .
\end{equation}
More general gluing procedures, called \emph{fusion}, can be performed when decorating a curve in the configuration with a non-minimal fibre and then gauging a subset of the classical flavour symmetry that rotates the hypermultiplets\footnote{The gauge-anomaly cancellation constraints the amount $N$ and the representation $\rho$ with which matter content charged under the gauge algebra $\mathfrak{g}$ appears, and thus the flavour symmetry $\mathfrak{f}$ is classically determined as $\surmL(N)$ if $\rho$ is \emph{complex}, as $\sormL(2N)$ if $\rho$ is \emph{pseudo-real}, as $\sprmL(N)$ if $\rho$ is \emph{real}.
    }. The price to pay is that one introduces an extra tensor (i.e.\ a new compact curve) that fits the requirement for anomaly cancellation and singularity of the Koidara classification, as in the following example:
\begin{equation}
    1 \ 2 \ \overset{\mathfrak{su}_2}{2} \ \overset{\mathfrak{su}_3}{2} [N_f=4] \text{ and } [N_f=4] \overset{\mathfrak{su}_3}{2} \ \overset{\mathfrak{su}_2}{2} \ 2 \ 1  
    \qquad \xrightarrow[]{\text{fusion}} \qquad 
    1 \ 2 \ \overset{\mathfrak{su}_2}{2} \ \overset{\mathfrak{su}_3}{2} \ {\color{blue}\overset{\mathfrak{su}_4}{2}} \  \overset{\mathfrak{su}_3}{2} \ \overset{\mathfrak{su}_2}{2} \ 2 \ 1 \ .
\end{equation}
Hence, the requirement that the self-intersection matrix $\eta$ has to have a one-dimensional null space automatically classifies the possible bases one can realise an LST with whilst employing the minimal blocks in \eqref{eqn:building_blocks}. In particular, calling $n_T$ the number of \emph{free-to-tune} tensor multiplets (i.e.\ the number of compact curves minus one that corresponds to the non-dynamical tensor responsible for the LST scale) in the LST, and $n_f$ the number of irreducible factors of its flavour symmetry $\mathfrak{f}$, the intersection matrix can be cast in the following form:
\begin{equation}
    \eta=\begin{pmatrix}
        \eta_{IJ} & \eta_{IA} \\
        \eta_{AI}&0
    \end{pmatrix} \ \text{with} \ I,J \in 1,\cdots , n_T+1 \text{ and } A \in 1,\cdots , n_f \,.
\end{equation}
This relates to the previous definition $\eta=C_i \cdot C_j$ by recalling that tensor multiplets (possibly) equipped with gauge algebras are compact curves $C_I$ in the geometry, whereas flavour symmetries can be interpreted as non-compact curves $C_A$ intersecting the corresponding compact curve on whose matter they act on.
Moreover, we also require that all the minors of $\eta$ must be negative defined, this condition can be physically understood as follows: \emph{cutting}\footnote{The procedure of \emph{cutting} is a tensor branch deformation that decompactifies one of the internal compact curve, this action realises two different decoupled theories localised along different points of the non-compact curve.} an LSTs curve configuration at any point must realise two SCFTs. Conversely,  when possible, an LST can be constructed by fusing two compatible SCFTs \cite{Lawrie:2023uiu}.

Although most of the models can be exhausted in this way, there exist a few extra low-rank LSTs that can be constructed from F-theory starting with a base space $\mathcal{B}$ of type
\begin{equation}
    P \quad , \quad I_0 \quad , \quad I \quad , \quad II \,, 
\end{equation}
where $P$ is a two-sphere, $I_0$ a torus, $I$ a compact curve with a node, and $II$ a compact curve with a cusp. All of which have trivial normal bundles. These models are comprised of a single curve, with self-intersection $0$ and finite size, that comes always paired with some gauge algebra for consistency\footnote{A full list of all the gauge algebras and matter content for a $0$-curve can be found in \cite{Bhardwaj:2015oru}.}. Such LSTs act as standalone models worth to be studied in depth.

The curve configuration of the F-theory engineering allows also to characterise the 2-Group structure constants \cite{DelZotto:2022ohj} appearing in \eqref{eqn:2-Group}. From the null eigenvector $\ell$ of the matrix $\eta$, multiplied by the smallest integer factor such that all its entries are integers, one can construct the following quantities: 
\begin{equation}
    \kappa_{F_A}=  \sum \limits_{i=1}^{N_T+1} \eta_{IA} \ell^I \ \quad , \quad \ \kappa_{R}= \sum \limits_{i=1}^{N_T+1} h^\vee_{II} \ell^I \quad , \quad \kappa_{\mathscr{P}}=  \sum \limits_{i=1}^{N_T+1}  (2-\eta_{II}) \ell^I \ ,
\end{equation} 
where $h^\vee_{II}$ is the dual Coxeter number of the gauge algebra equipped on the $I$-th compact curve in the base space. The index $A$ instead runs over all the non-compact curves $C_A$ in the base space contributing to the flavour symmetry factor $F_A$.

In this paper we focus on a simple family of F-theory models realised with a single curve of vanishing self-intersection. We analyse the specific case of a $P$ base space with an $I_N$ fibration that, depending on the fibre's monodromy \cite{Aspinwall:2000kf}, realises either an $\mathfrak{su}_N$ or an $\mathfrak{sp}_{\lfloor{\frac{N}{2}}\rfloor}$ gauge algebra. As usual in six dimensions, the matter content is enforced by gauge anomaly cancellation and permits only the following theories:
\begin{equation}\label{eqn:generic_models}
    [N_{f}= 16]\stackon{$0$}{$\mathfrak{su}_N$}[N_{\Lambda^2}= 2] \qquad \text{ and } \qquad  [N_{f}= 16]\stackon{$0$}{$\mathfrak{sp}_N$}[N_{\Lambda^2}= 1] \ .
\end{equation}
The null space of both these models is $\mathbb{Z}$; thus, their structure constants are:
\begin{equation} \label{eqn:structure-constants}
    \kappa_{R}= N \quad , \quad \ \kappa_{\mathscr{P}}= 2 \quad , \quad \kappa_{F_{f}}= 1 \quad , \quad \kappa_{F_{\Lambda^2}}= 1 \,,
\end{equation}
where with $F_\rho$ we indicated the flavour symmetry that rotates the matter in the representation $\rho$ of the gauge algebra.

There are also three outliers that need to be taken into account. For the algebras $\mathfrak{su}_6$ and $\mathfrak{sp}_3$, anomaly cancellation allows to trade of a second-rank anti-symmetric multiplet with a half third-rank anti-symmetric one plus a fundamental (or a fundamental and a half in the symplectic case), thus obtaining the following LST models:
\begin{equation}\label{eqn:Special_models}
    [N_{f}= 17]\stackunder{\stackon{$0$}{$\mathfrak{su}_6$}}{$[N_{\Lambda^2}= 1]$}[N_{\Lambda^3}= \tfrac12] \quad , \quad [N_{f}= 18]\stackon{$0$}{$\mathfrak{su}_6$}[N_{\Lambda^3}= 1] \quad , \quad  [N_{f}= 17+\tfrac{1}{2}]\stackon{$0$}{$\mathfrak{sp}_3$}[N_{\Lambda^3}= \tfrac12] \ .
\end{equation}
The structure constants related to the R-symmetry and the Poincar\'e symmetry for the models in \eqref{eqn:Special_models} are the same as the ones in \eqref{eqn:structure-constants}, as their value is independent of the matter content; whereas each flavour factor $F_\rho$ in \eqref{eqn:Special_models} rotating the matter in the $\rho$ representation, contributes as $1$ to the associated $\kappa_{F_\rho}$.
Hence, \eqref{eqn:generic_models} and \eqref{eqn:Special_models} complete the list of LSTs considered in this paper.

\section{Branes, magnetic quivers, and Hasse diagrams}
\label{sec:Branes-MQ}
Having introduced the little string theories \eqref{eqn:generic_models} and \eqref{eqn:Special_models}, we now turn to the Type IIA brane realisation, from which the magnetic quivers are constructed. Subsequently, brane dynamics, field theory, and magnetic quiver techniques are used to analyse the LST Higgs branches.

\subsection{Families of SU/Sp little string theories}
\begin{figure}[t]
    \centering
    \begin{subfigure}[t]{0.49\textwidth}
    \centering
\includegraphics[page=1]{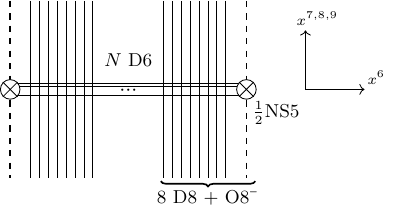}  
        \caption{}
        \label{subfig:branes_SU}
    \end{subfigure}
\begin{subfigure}[t]{0.49\textwidth}
\centering
\includegraphics[page=2]{pics/0-curve_figures.pdf}  
        \caption{}
        \label{subfig:branes_Sp}
    \end{subfigure}
    \caption{Brane realisation. \subref{subfig:branes_SU}: The $N$ D6 branes suspended between the two half NS5 branes carry an $\surm(N)$ gauge theory on their world-volume. F1 strings stretched between the D6 and 16 D8 branes lead to 16 fundamental hypermultiplets, while each half NS5 brane that is stuck on an O8 plane leads to a hypermultiplet transforming in the 2nd rank anti-symmetric of $\surm(N)$.  
    \subref{subfig:branes_Sp}: The $2N$ D6 branes are suspended between a single half NS5 brane and cross through the O8 plane; thus, the orientifold action leads to an $\sprm(N)$ gauge theory on the D6 world-volume. Since only one half NS5 brane is stuck on an O8 plane, there is a single anti-symmetric hypermultiplet.
    }
    \label{fig:branes_SU-Sp}
\end{figure}

The brane realisation is comprised of NS5, D6, D8 branes as well as O8 orientifold planes; see, for example, \cite{Hanany:1997gh,Hanany:1999sj} for an introduction. The space-time occupancy is indicated in Table \ref{tab:branes}. 
The brane systems for an $\surm(N)$ gauge theory with $N_f=16$ fundamental hypermultiplets and $N_{\Lambda^2}=2$ anti-symmetric hypermultiplets is shown in Figure \ref{subfig:branes_SU}. Similarly, the $\sprm(N)$ theory with $N_f=16$ and $N_{\Lambda^2}=1$ is realised via the brane configuration in Figure \ref{subfig:branes_Sp}.

\begin{table}[t]
\centering
\begin{tabular}{c|ccccccccccc}
 \toprule 
  & 0 & 1 & 2 & 3& 4 & 5 & 6  & 7 & 8 & 9 \\ \midrule
 NS5 & $\bullet$ & $\bullet$  & $\bullet$ & $\bullet$& $\bullet$ & 
$\bullet$&  \\
 D6 & $\bullet$ & $\bullet$ & $\bullet$ & $\bullet$ & $\bullet$ & 
$\bullet$ & $\bullet$
 \\
 D8/O$8^-$ & $\bullet$ & $\bullet$ & $\bullet$ & $\bullet$ & $\bullet$ & 
$\bullet$ & & $\bullet$ & $\bullet$ & $\bullet$
 \\ 
  \bottomrule 
\end{tabular}
\caption{Space time occupancy of branes. The $(5+1)$-dimensional world-volume theory is realised in $x^{0,1,\ldots,5}$. The D6 moduli along $x^{7,8,9}$ in between the D8 branes realise the Higgs branch directions.}
\label{tab:branes}
\end{table}

\subsubsection{Partial Higgsing via brane transitions}
\label{subsec:branes_generic}
The brane systems in Figure \ref{fig:branes_SU-Sp} allow us to deduce the possible partial Higgs mechanisms. To begin with, consider the $\surm(N)$ theory realised in Figure \ref{subfig:branes_SU}. For $N>3$, there are the following brane motions:
\begin{itemize}
    \item A single D6 brane can be split along the 16 D8 branes. The resulting 15 independent D6 segments can be moved off along the $x^{7,8,9}$ direction. This 15 (quaternionic) dimensional partial Higgs mechanism leaves behind a $\surm(N-1)$ gauge theory with $N_f=16$ fundamental and $N_{\Lambda^2}=2$ anti-symmetric hypermultiplets.
    
    \item If $N=2\ell$ even, a single half NS5 brane (still confined to a O8 plane) can move along $x^{7,8,9}$ such that the previously attached $N=2\ell$ D6 branes reconnect with their mirror images. After this (quaternionic) 1-dimensional transition, the residual world-volume theory is the $\sprm(\ell)$ theory of Figure \ref{subfig:branes_Sp}.

    However, there exists an alternative brane configuration for the same world-volume theory\footnote{The existence of two such brane systems was discussed in the context of 5-brane webs with O$7^-$ planes in \cite{Bergman:2015dpa}.}. Instead of moving just one stuck fractional NS5 along an O8, one can move up both NS5s on their respective orientifold. The residual brane system, after this apparent 2-dimensional transition, is $2\ell$ D6s intersecting two O$8^-$ orientifolds and 16 D8 branes. This is T-dual to $2\ell$ D5s inside a stack of 16 coincident D9 branes in the presence of an O$9^-$ plane \cite{Witten:1995gx,Douglas:1995bn} --- the brane realisation of $2\ell$ $\sorm(32)$-instantons on $\C^2$. Thus, this brane system is an alternative description of the $\sprm(\ell)$ world-volume theory. Below, it is demonstrated by other techniques that the transitions are in fact 2-dimensional.

    \item A subset of $2\ell$ D6 branes ($1\leq \ell<\lfloor \frac{N}{2} \rfloor$) can be moved along $x^{7,8,9}$, such that this stack of $2\ell$ D6s is no longer suspended between the fractional NS5 branes. This subsystem is composed of $2\ell$ D6s passing through two O$8^-$ planes as well as 16 D8 branes --- which is again the T-dual of $2\ell$ D5 branes in a stack of 16 coincident D9 brane in the presence of an O$9^-$ orientifold \cite{Witten:1995gx,Douglas:1995bn}.
    This fissions the brane system into two subsystems: the world-volume theory on the $2\ell$ D6s becomes an $\sprm(\ell)$ theory, due to the ADHM quiver; the remaining $N-2\ell$ D6s branes still realise a $\surm(N-2\ell)$ theory, as in Figure \ref{subfig:branes_SU}.
\end{itemize}
Analogously, the brane system in Figure \ref{subfig:branes_Sp} for the $\sprm(N)$ theory allows to deduce the following partial Higgs mechanisms
\begin{itemize}
    \item The $2N$ D6 branes split into a stack of $2(N-\ell)$ that are ending on one side on the fractional NS5 brane and a stack of $2\ell$ branes that are moved away from the NS5 along the $x^{7,8,9}$ direction (for $1\leq \ell \leq \lfloor \frac{N}{2}\rfloor$, $N>1$); the world-volume theory is the product $\sprm(N-\ell)\times\sprm(\ell)$ theory. Again, the latter subsystem stems from the T-dual of the instanton D$p$-D$(p+4)$ brane system. 

    Alternatively, one can start from the $\sprm(N)$ brane system without any fractional NS5 branes. Then, the fission of the stack of $2N$ D6 branes into a stack of $2(N-\ell)$ and a stack of $2\ell$ branes is just natural from the $N$ $\sorm(32)$-instanton point of view.
\end{itemize}
For the $\sprm(N)$ brane system this is the only type of transition.

\subsubsection{Partial Higgsing via branching rules}
\label{subsec:branching_generic}
The consistency of these partial Higgs mechanisms can be confirmed by validating the legitimacy of the symmetry breaking\footnote{The required branching rules can be conveniently analysed via \cite{Feger:2012bs,Feger:2019tvk,Yamatsu:2015npn}.}:
\begin{itemize}
    \item For any $N\geq 4$: \ul{$\surm(N)\to \surm(N-1)$}
    \begin{subequations}
    \begin{align}
    [1,0,\ldots,0]_{A_{N-1}} &\to   [1,0,\ldots,0]_{A_{N-2}} +  [0,0,\ldots,0]_{A_{N-2}} \\
    [0,1,0,\ldots,0]_{A_{N-1}} &\to  [0,1,0,\ldots,0]_{A_{N-2}} +  [1,0,\ldots,0]_{A_{N-2}} \\
     [1,0,\ldots,0,1]_{A_{N-1}} &\to     [1,0,\ldots,0,1]_{A_{N-2}} + [1,0,\ldots,0]_{A_{N-2}} + [0,\ldots,0,1]_{A_{N-2}}  \\ 
     &\qquad + [0,0,\ldots,0]_{A_{N-2}}  \notag
    \end{align}
    \label{eq:decomp_SUN-SUN-1}%
    \end{subequations}
such that the $\surm(N-1)$ fundamentals appearing from the adjoint can be compensated by the fundamentals appearing in the decomposition of $\Lambda^2$. Thus, $N_f$ remains $16$ and the number of (quaternionic) singlets is indeed $16-1=15$. The one singlet appearing in the adjoint signals that the commutant of $\surm(N-1)$ inside $\surm(N)$ is a rank 1 group, which is a $\urm(1)$. Thus, the transition is an $a_{15}$. 

For $N=3$, the logic is essentially the same as in \eqref{eq:decomp_SUN-SUN-1}, but $\Lambda^2$ coincides with the anti-fundamental representation. Therefore, two more singlets appear in the decomposition of the matter fields and the transition is an $a_{17}$ instead. 

\item For $N=2\ell$ even: \ul{$\surm(2\ell) \to \sprm(\ell)$}
\begin{subequations}
   \begin{align}
    [1,0,\ldots,0]_{A_{N-1}} &\to   [1,0,\ldots,0]_{C_{\ell}} \\
    [0,1,0,\ldots,0]_{A_{N-1}} &\to  [0,1,\ldots,0]_{C_{\ell}} +  [0,0,\ldots,0]_{C_{\ell}} \\
     [1,0,\ldots,0,1]_{A_{N-1}} &\to     [2,0,\ldots,0]_{C_{\ell}} + [0,1,0\ldots,0]_{C_{\ell}}   
    \end{align}
    \end{subequations}
    such that one of the $\sprm(\ell)$ anti-symmetric hypermultiplets needs to compensate for the anti-symmetric appearing in the decomposition of the adjoint, leaving behind a single anti-symmetric hypermultiplet. Moreover, there are two singlets appearing from the matter fields, consistent with a 2-dimensional transition. As there is no singlet in the decomposition of the adjoint, the commutant of $\sprm(\ell)$ inside $\surm(2\ell)$ is at most a discrete group. The transition is $h_{2,\ell}$, see below 

\item For any $N$ and $1\leq \ell < \lfloor \frac{N}{2}\rfloor$: \ul{$\surm(N) \to \sprm(\ell) \times \surm(N-2\ell)$}. To analyse this branching, first consider $\surm(N) \to \surm(2\ell) \times \surm(N-2\ell)$
\begin{subequations}
\begin{align}
 [1,0,\ldots,0]_{A_{N-1}} &\to   
 [1,0,\ldots,0]_{A_{2\ell-1}} \otimes [0,\ldots,0]_{A_{N-2\ell-1}}
  \\ &\qquad +  
 [0,\ldots,0]_{A_{2\ell-1}} \otimes
 [0,\ldots,0,1]_{A_{N-2\ell-1}}  \notag \\
    [0,1,0\ldots,0]_{A_{N-1}} &\to  
    [0,1,0\ldots,0]_{A_{2\ell-1}} \otimes [0,\ldots,0]_{A_{N-2\ell-1}}  \\
   &\qquad + 
    [0,\ldots,0]_{A_{2\ell-1}} \otimes
    [0,1,0,\ldots,0]_{A_{N-2\ell-1}} \notag \\
    &\qquad +  [1,0,\ldots,0]_{A_{2\ell-1}} \otimes [1,0,\ldots,0]_{A_{N-2\ell-1}} \notag \\
     [1,0,\ldots,0,1]_{A_{N-1}} &\to     
     [1,0,\ldots,0,1]_{A_{2\ell-1}}
     \otimes [0,\ldots,0]_{A_{N-2\ell-1}} \\
      &\qquad + 
    [0,\ldots,0]_{A_{2\ell-1}} \otimes [1,0,\ldots,0,1]_{A_{N-2\ell-1}}  \notag \\
    &\qquad
     + [1,0,\ldots,0]_{A_{2\ell-1}} \otimes [0,\ldots,0,1]_{A_{N-2\ell-1}} \notag \\
    &\qquad
     + [0,\ldots,0,1]_{A_{2\ell-1}} \otimes [1,0,\ldots,0]_{A_{N-2\ell-1}} \notag \\
    &\qquad
     + [0,\ldots,0]_{A_{2\ell-1}} \otimes [0,0,\ldots,0]_{A_{N-2\ell-1}}  \notag 
\end{align}
\end{subequations}
because the appearing bifundamentals in the decomposition of the adjoint cannot be compensated, it is clear that this breaking is not a consistent partial Higgs mechanism. However, the further breaking $\surm(2\ell) \to \sprm(\ell)$ is still legitimate. Combining both, one finds for $\surm(N)\to \sprm(\ell) \times \surm(N-2\ell)$
\begin{subequations}
\begin{align}
 [1,0,\ldots,0]_{A_{N-1}} &\to   [1,0,\ldots,0]_{C_{\ell}} \otimes [0,\ldots,0]_{A_{N-2\ell-1}} 
 \\
 &\qquad + 
 [0,\ldots,0]_{C_{\ell}} \otimes
 [0,\ldots,0,1]_{A_{N-2\ell-1}}  \notag \\
    [0,1,0,\ldots,0]_{A_{N-1}} &\to  
    [0,1,0,\ldots,0]_{C_{\ell}} \otimes [0,\ldots,0]_{A_{N-2\ell-1}} \\
    &\qquad+ [0,\ldots,0]_{C_{\ell}} \otimes [0,\ldots,0]_{A_{N-2\ell-1}} \notag \\
    &\qquad + [0,\ldots,0]_{C_{\ell}} \otimes [0,1,\ldots,0]_{A_{N-2\ell-1}} \notag \\
    &\qquad +  [1,0,\ldots,0]_{C_{\ell}} \otimes [1,0,\ldots,0]_{A_{N-2\ell-1}} \notag \\
     [1,0,\ldots,0,1]_{A_{N-1}} &\to     
     [2,0,\ldots,0]_{C_{\ell}} \otimes [0,\ldots,0]_{A_{N-2\ell-1}}  \\
    &\qquad
     + [0,1,0,\ldots,0]_{C_{\ell}} \otimes [0,\ldots,0]_{A_{N-2\ell-1}} \notag \\
    &\qquad
     + [0,\ldots,0]_{C_{\ell}} \otimes [1,0,\ldots,0,1]_{A_{N-2\ell-1}} \notag \\
     &\qquad + [1,0,\ldots,0]_{C_{\ell}} \otimes [0,\ldots,0,1]_{A_{N-2\ell-1}} \notag \\
     &\qquad
     + [1,0,\ldots,0]_{C_{\ell}} \otimes [1,0,\ldots,0]_{A_{N-2\ell-1}} \notag \\
     &\qquad
     + [0,\ldots,0]_{C_{\ell}} \otimes [0,\ldots,0]_{A_{N-2\ell-1}}   \notag
\end{align}
\end{subequations}
which is a consistent partial Higgs mechanism. There is one singlet in the decomposition of the adjoint, indicating a commutant of rank 1. Moreover, there are two singlets from the matter fields; hence the transition is of quaternionic dimension 1. 
\item For $N>1$ and $1\leq \ell \leq \lfloor N\rfloor$: \ul{$\sprm(N) \to \sprm(N-\ell) \times \sprm(\ell)$}.
\begin{subequations}
\begin{align}
 [1,0,\ldots,0]_{C_{N}} &\to   
 [1,0,\ldots,0]_{C_{N-\ell}}  \otimes [0,\ldots,0]_{C_{\ell}}
 + [0,\ldots,0]_{C_{N-\ell}}  \otimes [1,0,\ldots,0]_{C_{\ell}} \\
    [0,1,0\ldots,0]_{C_{N}} &\to  
    [0,1,0\ldots,0]_{C_{N-\ell}} \otimes [0,\ldots,0]_{C_{\ell}}
    + [0,\ldots,0]_{C_{N-\ell}}  \otimes [0,1,0,\ldots,0]_{C_{\ell}} \\
    &\qquad +  [1,0,\ldots,0]_{C_{N-\ell}} \otimes [1,0,\ldots,0]_{C_{\ell}} 
    +  [0,\ldots,0]_{C_{N-\ell}} \otimes [0,\ldots,0]_{C_{\ell}} \notag \\
     [2,0,\ldots,0]_{C_{N}} &\to     
     [2,0,\ldots,0]_{C_{N-\ell}} \otimes [0,\ldots,0]_{C_{\ell}} + [0,\ldots,0]_{C_{N-\ell}}  \otimes [2,0,\ldots,0]_{C_{\ell}}  \\
    &\qquad
     + [1,0,\ldots,0]_{C_{N-\ell}} \otimes [1,0,\ldots,0]_{C_{\ell}}  \notag 
\end{align}
\end{subequations}
which shows consistency with the splitting in the brane system. The appearing singlet indicates a $1$-dimensional transition. As there is no singlet in the decomposition of the adjoint, the corresponding slice is not expected to have a global symmetry.

Note that this analysis is essentially the statement that $N$ $\sorm(32)$-instantons can split into $N-\ell$ and $\ell$ $\sorm(32)$-instantons. To see this, note that the 6d effective description matches the $D$-type ADHM quiver. A related work can be found in \cite{Bourget:2022ehw}.
\end{itemize}

\begin{figure}[ht]
    \centering
\includegraphics[page=10,scale=0.87]{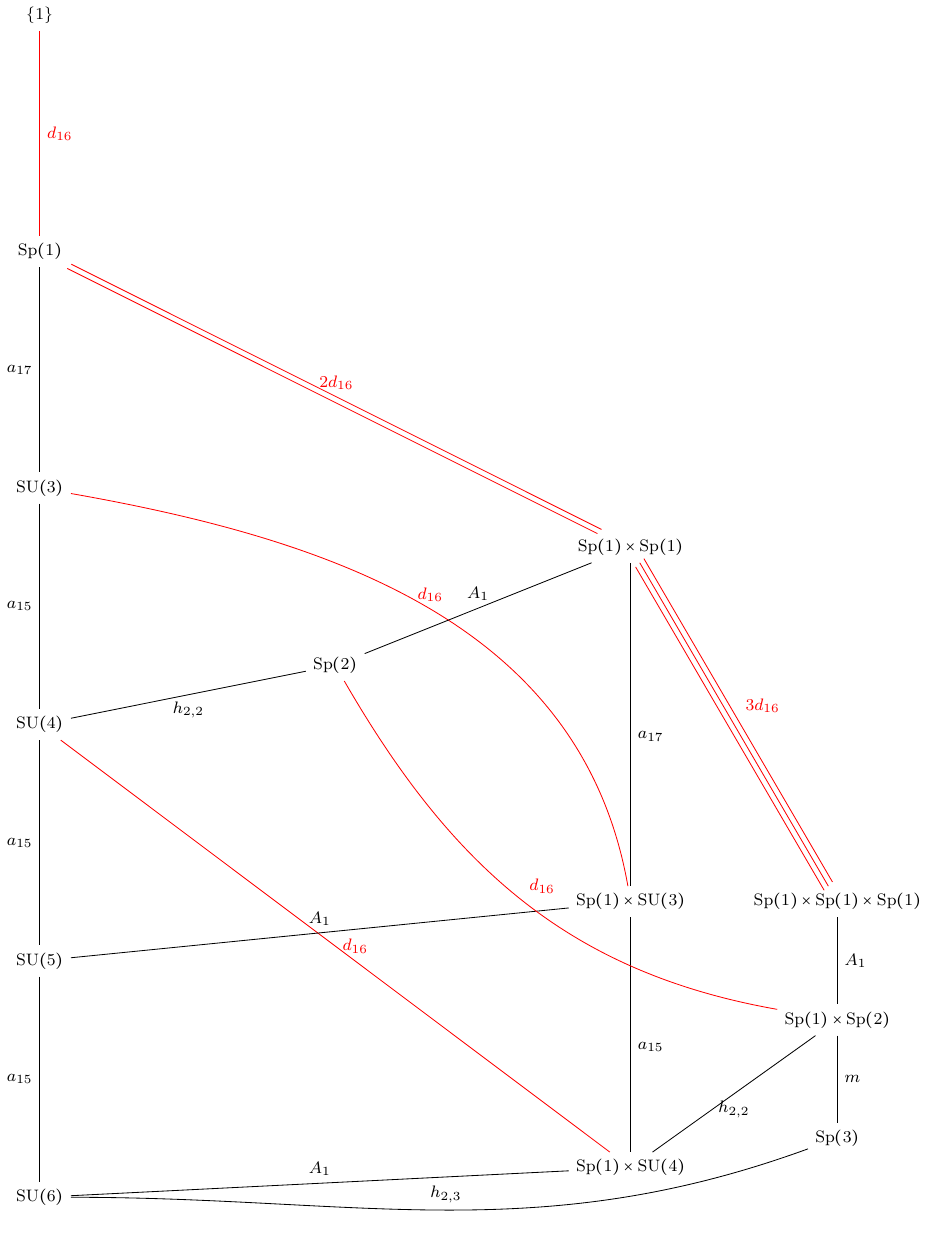}
    \caption{Hasse diagram of $\surm(6)$, $N_f=16$, $N_{\Lambda^2}=2$ via partial Higgs mechanism.}
    \label{fig:Hasse_SU6_Higgs}
\end{figure}

In summary, based on brane motions and field theory arguments, one can establish a phase diagrams for Higgs branch RG-flows, as shown in Figures \ref{subfig:higgs_SU4},  \ref{subfig:higgs_SU5}, \ref{fig:Hasse_SU6_Higgs} for the cases of an $\surm(N)$ theory with $N=4,5,6$. Below, these diagrams are supplemented by the geometric information of each minimal RG-flow.

\subsubsection{Higgs branch Hasse diagram via quiver subtraction}
\label{subsec:QuiverSub_generic}

\begin{sidewaysfigure}[h]
    \centering
    \begin{subfigure}[t]{0.33\textwidth}
    \flushleft
  \includegraphics[page=5,scale=0.9]{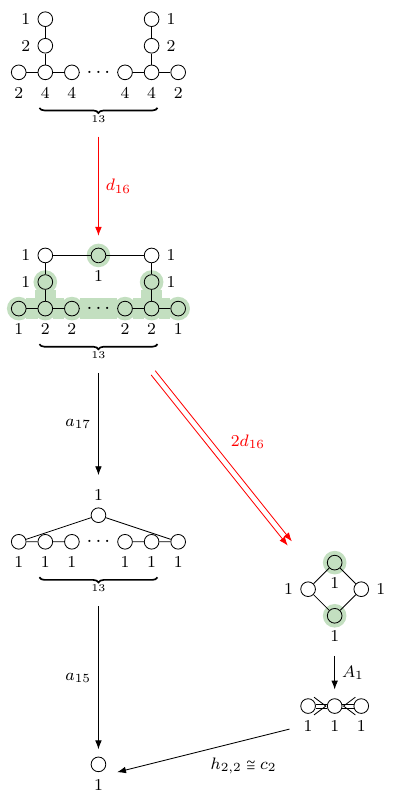}
  \caption{Quiver subtraction}
  \label{subfig:MQ_sub_SU4}
    \end{subfigure}
    \hfill
     \begin{subfigure}[t]{0.33\textwidth}
    \centering
  \includegraphics[page=17,scale=0.9]{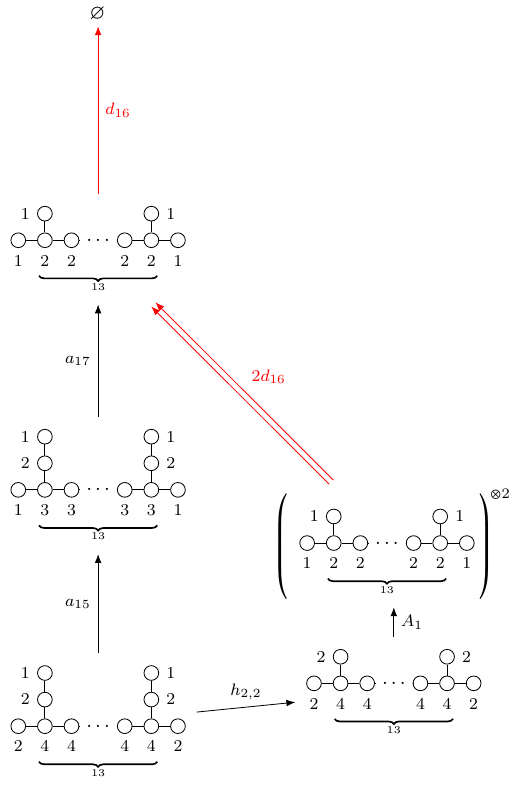}
  \caption{Decay and fission}
  \label{subfig:Decay_fission_SU4}
    \end{subfigure}
    \hfill
    \begin{subfigure}[t]{0.33\textwidth}
    \centering
  \includegraphics[page=6,scale=0.9]{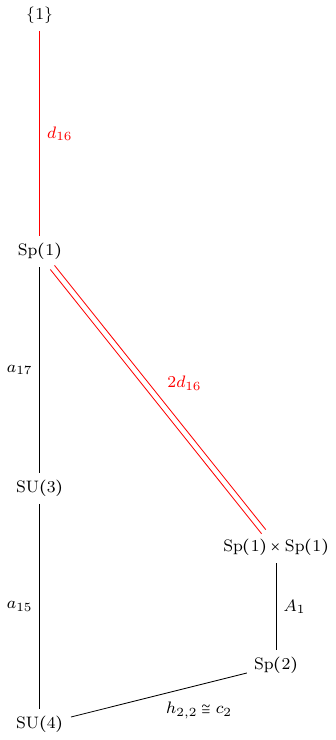}
  \caption{RG-flows}
    \label{subfig:higgs_SU4}
    \end{subfigure}
    \caption{Hasse diagram of $\surm(4)$, $N_f=16$, $N_{\Lambda^2}=2$. \subref{subfig:MQ_sub_SU4} via quiver subtraction. Green background shading indicates ``decorations''. 
    \subref{subfig:Decay_fission_SU4} via decay and fission algorithm. \subref{subfig:higgs_SU4} via partial Higgs mechanism.}
    \label{fig:Hasse_SU4}
\end{sidewaysfigure}

\begin{sidewaysfigure}[h]\hspace*{-100pt}
    \centering
    \begin{subfigure}[t]{0.4\textwidth}
    \centering
  \includegraphics[page=7,scale=0.75]{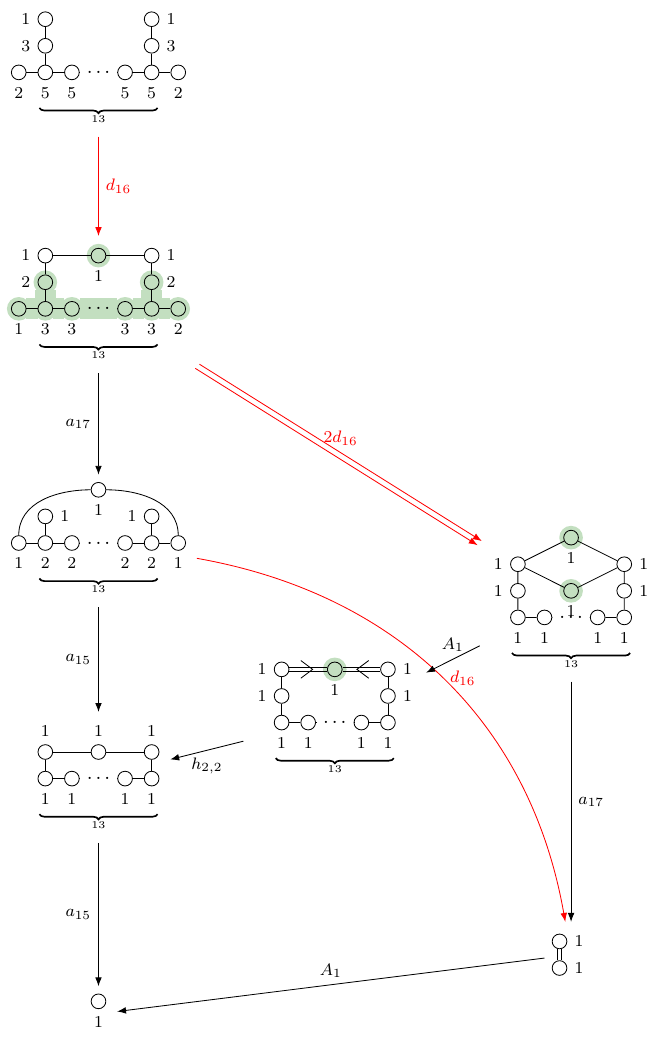}
  \caption{Quiver subtraction}
  \label{subfig:MQ_sub_SU5}
    \end{subfigure}
     \begin{subfigure}[t]{0.4\textwidth}
    \centering
  \includegraphics[page=18,scale=0.75]{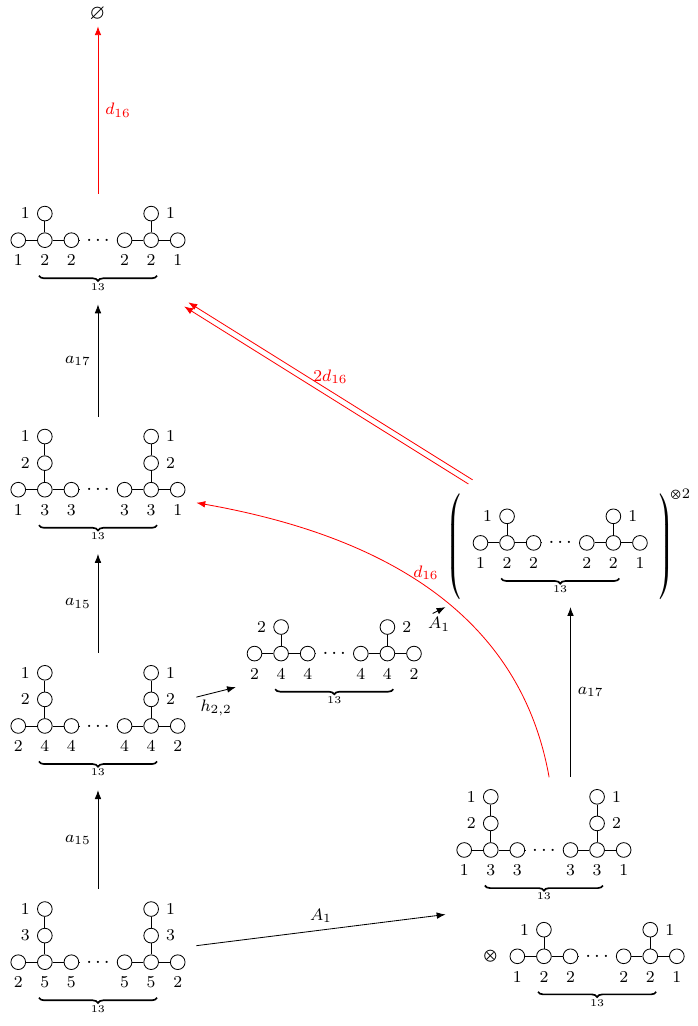}
  \caption{Decay and fission}
  \label{subfig:Decay_fission_SU5}
    \end{subfigure}
    \begin{subfigure}[t]{0.1\textwidth}
    \centering
  \includegraphics[page=8,scale=0.75]{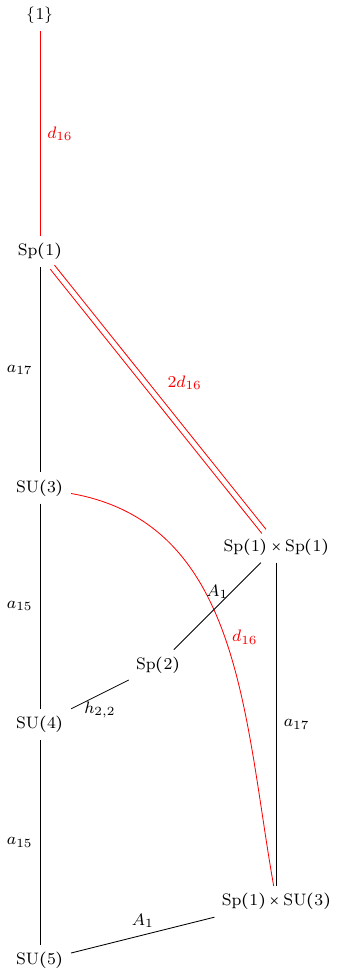}
  \caption{RG-flows}
    \label{subfig:higgs_SU5}
    \end{subfigure}
    \caption{Hasse diagram of $\surm(5)$, $N_f=16$, $N_{\Lambda^2}=2$. \subref{subfig:MQ_sub_SU5} via quiver subtraction. Green background shading indicates ``decorations''. 
    \subref{subfig:Decay_fission_SU5} via decay and fission algorithm.
    \subref{subfig:higgs_SU5} via partial Higgs mechanism.}
    \label{fig:Hasse_SU5}
\end{sidewaysfigure}

\begin{figure}[ht]
    \centering
\includegraphics[page=9,scale=0.87]{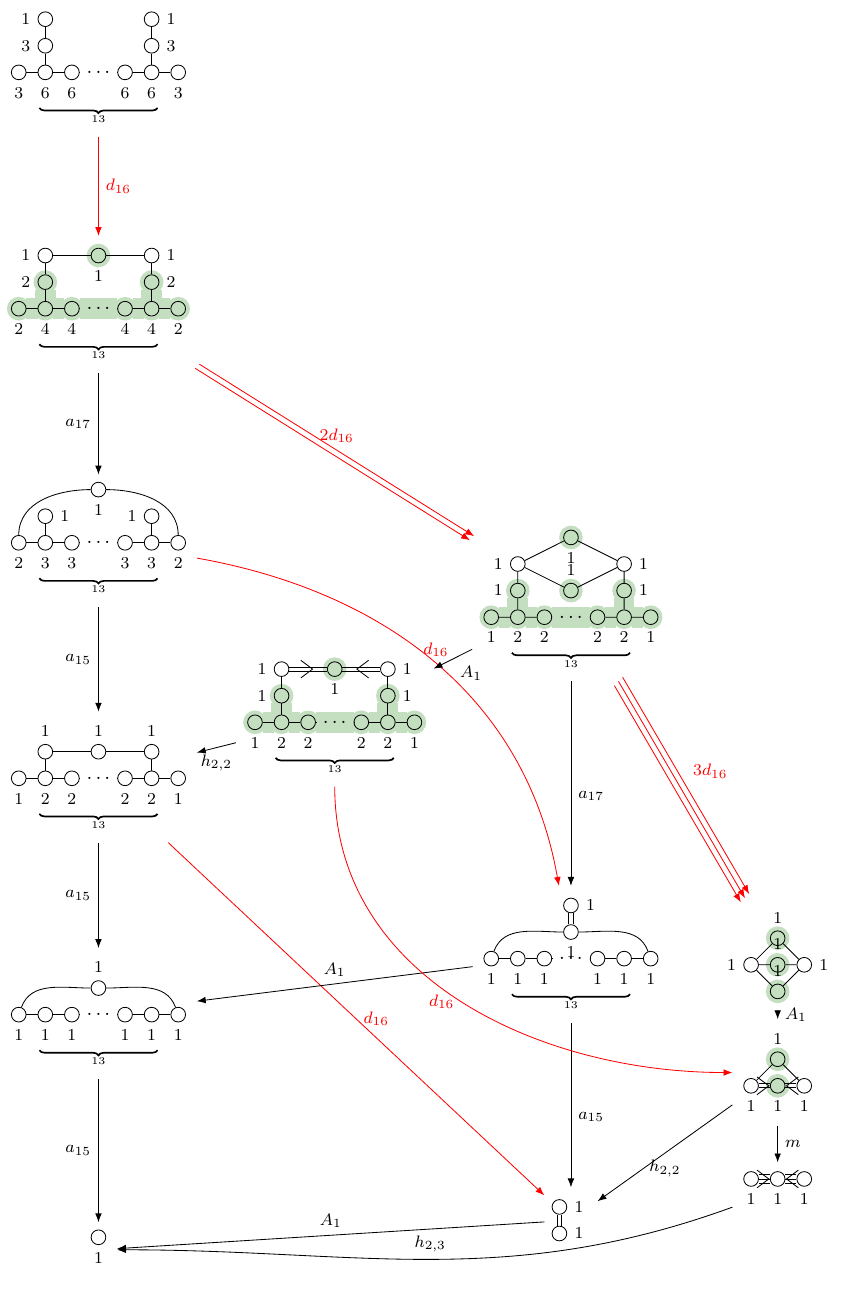}
    \caption{Hasse diagram of $\surm(6)$, $N_f=16$, $N_{\Lambda^2}=2$ via quiver subtraction. Green background shading indicates ``decorations''.}
    \label{fig:MQ_sub_SU6}
\end{figure}

Starting from the brane realisations of Figure \ref{fig:branes_SU-Sp}, the magnetic quivers are derived straightforwardly \cite{Cabrera:2019izd}. Concretely, one finds
\begin{itemize}
    \item $\surm(N)$ with $N=2\ell$ even
    \begin{align}
\raisebox{-.5\height}{
 \includegraphics[page=3]{pics/0-curve_figures.pdf}
 }
 \label{eq:MQ_SU-even}
\end{align}
    \item $\surm(N)$ with $N=2\ell+1$ odd
        \begin{align}
\raisebox{-.5\height}{
 \includegraphics[page=4]{pics/0-curve_figures.pdf}
 }
  \label{eq:MQ_SU-odd}
\end{align}
\end{itemize}
Building on these, the Higgs branch Hasse diagram is derived via (standard) quiver subtraction \cite{Bourget:2019aer,Bourget:2020mez,Bourget:2022ehw,Bourget:2022tmw} (including ``decorations''). Figures \ref{subfig:MQ_sub_SU4}, \ref{subfig:MQ_sub_SU5}, \ref{fig:MQ_sub_SU6}, provide illustrative examples for $\surm(N)$ theories with $N=4,5,6$.

Similarly, for the $\sprm(\ell)$ theories on a single $0$-curve, the magnetic quiver is given by
  \begin{align}
\raisebox{-.5\height}{
 \includegraphics[page=11]{pics/0-curve_figures.pdf}
 }
  \label{eq:MQ_Sp}
\end{align}
which is a ``bad'' quiver gauge theory, in the sense of \cite{Gaiotto:2008ak}. While this means that most computational approaches may diverge, one can nonetheless apply algorithms like \emph{decay and fission} \cite{Bourget:2023dkj,InvQuiverSubHiggs} to analyse the Higgs branch RG-flows of the $\sprm$-type LSTs in \eqref{eqn:generic_models}. As shown below, \eqref{eq:MQ_Sp} is formally a stack of $\ell$ $D_{16}$ Dynkin quivers and the fission of the magnetic quiver yields valuable insight into the Higgs branch RG-flows of the LST. 

It does remain an open challenge to provide a ``good'' magnetic quiver that can reproduce the same features as \eqref{eq:MQ_Sp}; a related issue has been encountered in \cite{DelZotto:2023nrb}.

\begin{figure}[ht]
    \centering
\includegraphics[page=19,scale=0.95]{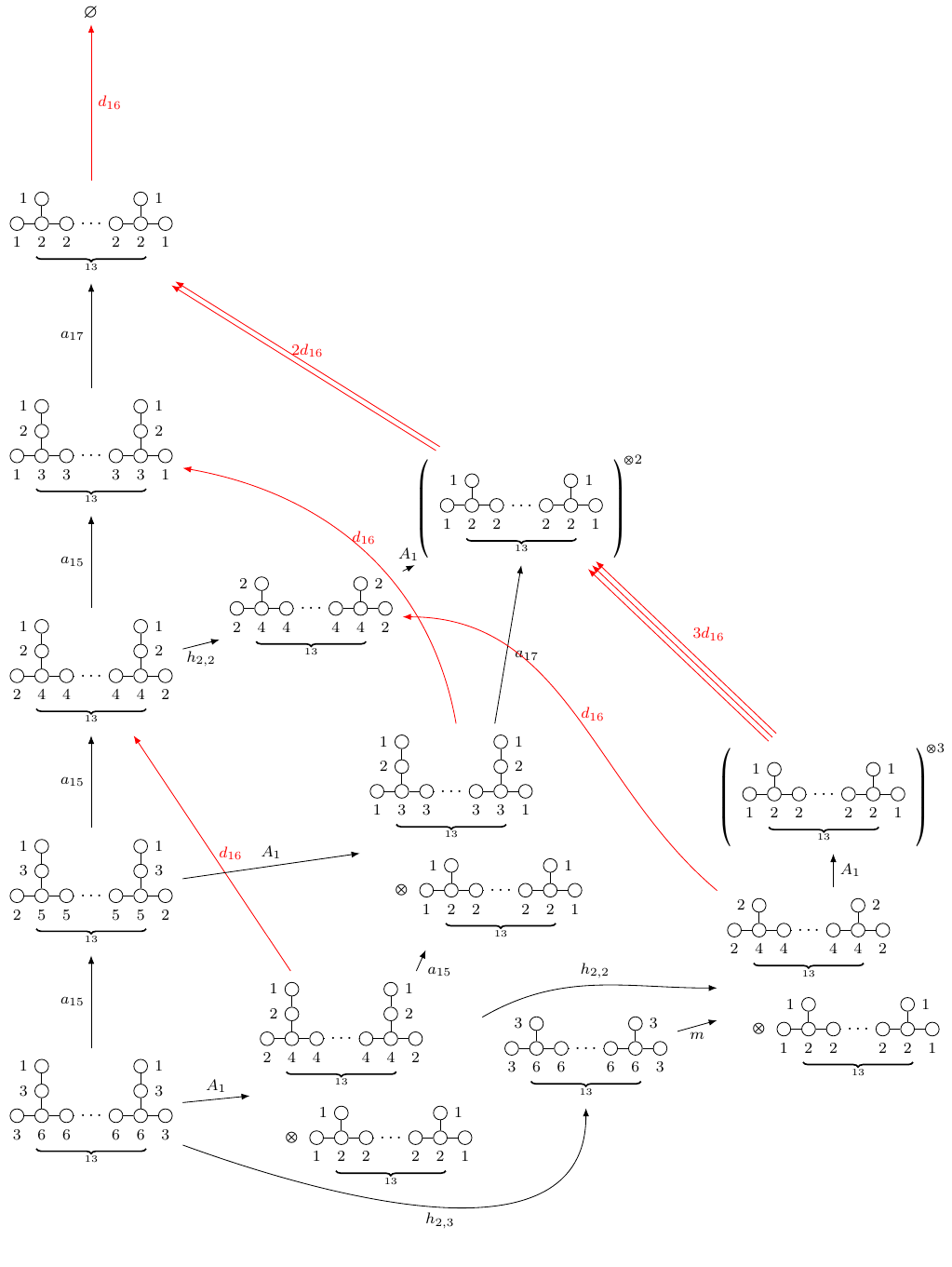}
    \caption{Hasse diagram of $\surm(6)$, $N_f=16$, $N_{\Lambda^2}=2$ via the decay and fission algorithm.}
    \label{fig:Decay_fission_SU6}
\end{figure}

\paragraph{General transitions.}
The quiver subtraction algorithm --- introduced in \cite{Bourget:2019aer} and further developed in \cite{Bourget:2020mez,Bourget:2022ehw,Bourget:2022tmw}  --- allows to deduce the nature of the partial Higgs transitions. Fairly straightforward are the $a_{15}$, $a_{17}$, and $d_{16}$ transitions. However, as illustrated in the examples in Figures \ref{subfig:MQ_sub_SU4}, \ref{subfig:MQ_sub_SU5}, and \ref{fig:MQ_sub_SU6}, there are a collection of other transitions. These can be systematically analysed as follows:   
\begin{itemize}
    \item \ul{$\surm(2\ell)\to\sprm(\ell)$}: Starting from \eqref{eq:MQ_SU-even}, subtracting $\ell$ consecutive $d_{16}$ transitions leads to 
        \begin{align}
\raisebox{-.5\height}{
 \includegraphics[page=12]{pics/0-curve_figures.pdf}
 } \label{eq:MQ_Spl}
\end{align}
which is composed of $\ell$ decorated $\urm(1)$ nodes (in green). Physically, this corresponds to the symplectic leaf $\surm(2\ell) \to (\sprm(1))^{\otimes \ell}$. After collapsing all the (green) decorated nodes, one finds
\begin{align}
\raisebox{-.5\height}{
 \includegraphics[page=13]{pics/0-curve_figures.pdf}
 } 
 \label{eq:MQ_leaf_SU2Sp}
\end{align}
which describes the symplectic leaf $\surm(2\ell) \to \sprm(\ell)$. The magnetic quiver \eqref{eq:MQ_leaf_SU2Sp} (with non-simply laced edges of order $\ell$) is known to describe the $h_{2,\ell}$ slice \cite{Bourget:2020asf,Bourget:2021siw}. 

\item \ul{$\surm(2\ell)\to\sprm(k) \times \surm(2(\ell-k))$}: Again, starting from \eqref{eq:MQ_SU-even}, subtracting  $k$ consecutive $d_{16}$ transitions ($1\leq k <\ell$) leads to the magnetic quiver for the symplectic leaf $\surm(2\ell) \to (\sprm(1))^{\otimes k}$. Further subtracting one $a_{17}$ and $(2(\ell-k)-1)$ consecutive $a_{15}$ transitions, leads to the magnetic quiver with $k$ decorated nodes (in green)
\begin{align}
\raisebox{-.5\height}{
 \includegraphics[page=14]{pics/0-curve_figures.pdf}
 } \label{eq:MQ_kxSp1}
\end{align}
which describes the symplectic leaf $\surm(2\ell) \to \surm(2(\ell-k))  \times (\sprm(1))^{\otimes k}$. Collapsing all the (green) decorated $\urm(1)$ nodes results in
\begin{align}
\raisebox{-.5\height}{
 \includegraphics[page=15]{pics/0-curve_figures.pdf}
 }\label{eq:MQ_Spk}
\end{align}
which is the sought after magnetic quiver for the symplectic leaf $\surm(2\ell) \to \surm(2(\ell-k)) \times \sprm(k)$. Ungauging on the long node in \eqref{eq:MQ_Spk} then results in the quiver for the $A_{1}$ singularity \cite{Bourget:2022tmw}.

\item \ul{$\surm(2\ell+1)\to\sprm(k) \times \surm(2(\ell-k)+1)$}: here, analogous arguments apply. After $k$ consecutive $d_{16}$ transitions, one performs one $a_{17}$ transitions, followed by $2(\ell-k)$ $a_{15}$ transitions. The resulting magnetic quiver is \eqref{eq:MQ_kxSp1} with $k$ decorated $\urm(1)$ nodes. After collapsing these nodes, the magnetic quiver for the symplectic leaf $\surm(2\ell+1)\to\sprm(k) \times \surm(2(\ell-k)+1) $ is \eqref{eq:MQ_Spk}; thus, leading to a $A_{1}$ singularity.

\item \ul{$\sprm(\ell)\to \sprm(\ell-k) \times \sprm(k)$}: Performing $\ell$ consecutive $d_{16}$ transitions on \eqref{eq:MQ_SU-even} leads to \eqref{eq:MQ_Spk}. From the $k$ decorated $\urm(1)$, one merges $l$ as well as $k$ of them, which results in the magnetic quiver for the symplectic leaf $\surm(2\ell) \to \sprm(\ell-k) \times \sprm(k)$. To understand the transition $\sprm(\ell) \to \sprm(\ell-k) \times \sprm(k)$, one only needs to focus on the collapse of the last two decorated nodes (with non-simply laced edges)
\begin{align}
\raisebox{-.5\height}{
 \includegraphics[page=16]{pics/0-curve_figures.pdf}
 }\label{eq:MQ_SpkxSpl}
\end{align}
which is analysed by the same arguments as in \cite{Bourget:2022tmw}.
\end{itemize}

\subsubsection{Higgs branch Hasse diagram via decay and fission}
Given a magnetic quiver, using the \emph{decay and fission} algorithm \cite{Bourget:2023dkj,InvQuiverSubHiggs} allows us to deduce possible Higgs branch RG-flows of the theory in question. For the case of the $\surm(N)$ theories with $N=4,5,6$, the results of the decay and fission algorithm are shown in Figures \ref{subfig:Decay_fission_SU4}, \ref{subfig:Decay_fission_SU5}, and \ref{fig:Decay_fission_SU6}. This offers a consistence check of the Higgsings deduced from branching rules and brane transitions. Moreover, it provides validation that the derived magnetic quivers capture the Higgs branch of electric theory sufficiently well.

\begin{itemize}

\item \ul{$\surm(2\ell)\to\surm(2\ell-1)$}: Starting from \eqref{eq:MQ_SU-even}, one observes a linear chain of $15$ balanced gauge nodes. This induces the following $a_{15}$ decay \cite{Bourget:2023dkj,InvQuiverSubHiggs}:
    \begin{align}
\raisebox{-.5\height}{
 \includegraphics[page=1]{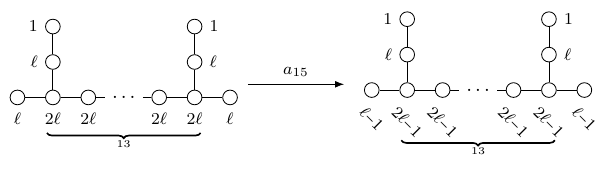}
 }
\end{align}
and the resulting magnetic quiver is recognised as that of $\surm(2\ell-1)$ on a 0-curve, cf.\ \eqref{eq:MQ_SU-odd}. 

\item \ul{$\surm(2\ell)\to\sprm(k) \times \surm(2(\ell-k))$}: Again, starting from \eqref{eq:MQ_SU-even}, one recognises a stack of $\ell$ affine $d_{16}$ Dynkin diagrams. Such a stack can be fission into any partition $(\ell-k,k)$ for $k\in \{1,\ldots,\ell-1\}$ \cite{Bourget:2023dkj}. In terms of quivers, this becomes
\begin{align}
\raisebox{-.5\height}{
 \includegraphics[page=2]{pics/0-curve_IQS_figures.pdf}
 }
\end{align}
and one finds the magnetic quivers for an $\surm(2\ell-2k)$ theory on a 0-curve \eqref{eq:MQ_SU-even} and the magnetic quiver for an $\sprm(k)$ theory on a 0-curve \eqref{eq:MQ_Sp}. The transition geometry is detailed in \cite{InvQuiverSubHiggs}.

\item \ul{$\surm(2\ell)\to\sprm(\ell)$}: Starting from \eqref{eq:MQ_SU-even}, the decay algorithm yield a transition of the form
    \begin{align}
\raisebox{-.5\height}{
 \includegraphics[page=4]{pics/0-curve_IQS_figures.pdf}
 }
\end{align}
wherein the slice is computed to be $h_{2,l}$ \cite{InvQuiverSubHiggs}.
The resulting magnetic quiver is, in fact, that of the $\sprm(\ell)$ theory \eqref{eq:MQ_Sp}.

\item \ul{$\surm(2\ell+1)\to \surm(2\ell)$}: For $\ell >1$, there is a linear chain of $15$ balanced gauge nodes in the magnetic quiver \eqref{eq:MQ_SU-odd}; thus, implies the following decay
 \begin{align}
\raisebox{-.5\height}{
 \includegraphics[page=5]{pics/0-curve_IQS_figures.pdf}
 }
\end{align}
which produces the magnetic quiver \eqref{eq:MQ_SU-even} for the $\surm(2\ell)$ theory on a 0-curve.

For $\ell=1$, the quiver \eqref{eq:MQ_SU-odd} contains a linear chain of $17$ balanced gauge nodes; Therefore, indicating an $a_{17}$ decay
\begin{align}
\raisebox{-.5\height}{
 \includegraphics[page=6]{pics/0-curve_IQS_figures.pdf}
 }
\end{align}
and the resulting magnetic quiver is that of the $\surm(2)\cong\sprm(1)$ theory in a 0-curve, cf.\ \eqref{eq:MQ_Sp}.

\item \ul{$\surm(2\ell+1)\to\sprm(k) \times \surm(2(\ell-k)+1)$}: The magnetic quiver \eqref{eq:MQ_SU-odd} contains a stack of $\ell$ affine $d_{16}$ Dynkin diagrams, which can fission in any partition $(\ell-k,k)$ with $k\in \{1,2,\ldots,\ell\}$ by \cite{Bourget:2023dkj,InvQuiverSubHiggs}. The quiver fission takes the form 
 \begin{align}
\raisebox{-.5\height}{
 \includegraphics[page=7]{pics/0-curve_IQS_figures.pdf}
 }
\end{align}
and one recovers the magnetic quivers for the $\surm(2\ell-2k+1)$ theory \eqref{eq:MQ_SU-odd} and the $\sprm(k)$ theory \eqref{eq:MQ_Sp}, both defined on a 0-curve.

\item \ul{$\sprm(\ell)\to \sprm(\ell-k) \times \sprm(k)$}: Starting from the formal magnetic quiver \eqref{eq:MQ_Sp}, the obvious stack of $\ell$ affine $d_{16}$ Dynkin diagrams can fission in any partition $(\ell-k,k)$ for $k\in \{1,2,\ldots,\ell-1\}$. In terms of quivers, this becomes
 \begin{align}
\raisebox{-.5\height}{
 \includegraphics[page=8]{pics/0-curve_IQS_figures.pdf}
 }
\end{align}
and one finds the magnetic quivers for the $\sprm(\ell-k)$ theory and the $\sprm(k)$ theory, both defined on a 0-curve.
\end{itemize}
In summary, the \emph{decay and fission} algorithm recovers the same Higgs branch RG-flows as predicted by brane motions in Section~\ref{subsec:branes_generic} as well as branching rules in Section~\ref{subsec:branching_generic}. Moreover, it generates the same Higgs branch Hasse diagram as derived via quiver subtraction in Section~\ref{subsec:QuiverSub_generic}.

\FloatBarrier

\subsection{Special SU/Sp little string theories}
\label{sec:special_theories}
\begin{figure}[h]
    \centering
    \begin{subfigure}[t]{0.49\textwidth}
    \centering
\includegraphics[page=35]{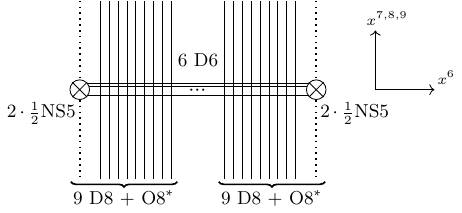}  
        \caption{$\surm(6)^{\prime\prime}$}
        \label{subfig:branes_SU6_Nf=18}
    \end{subfigure}
    \begin{subfigure}[t]{0.49\textwidth}
\centering
\includegraphics[page=36]{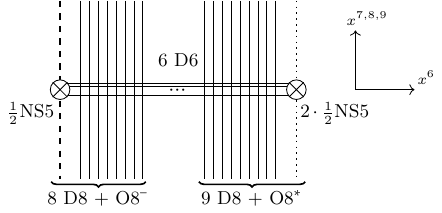}  
        \caption{$\surm(6)^{\prime}$}
        \label{subfig:branes_SU6_Nf=17}
    \end{subfigure}
\begin{subfigure}[t]{0.49\textwidth}
\centering
\includegraphics[page=37]{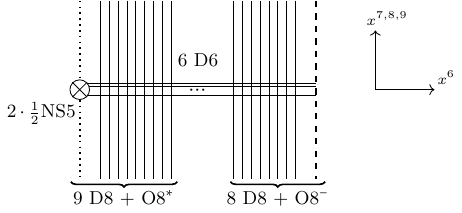}  
        \caption{$\sprm(3)^{\prime}$}
        \label{subfig:branes_Sp3_Nf=17-5}
    \end{subfigure}
    \caption{Brane realisation of the special SU/Sp theories. \subref{subfig:branes_SU6_Nf=18}: a stack of 6 D6 branes in between two O8${}^\ast$s; as in \cite{Cabrera:2019izd}, there are two stuck half NS5s on each O8${}^\ast$ and 9 D8s parallel to it. As the D6s are between NS5s, there is an $\surm(6)$ vector. The $18$ D8s contribute $N_f=18$ fundamentals and each NS5-O8${}^\ast$ boundary introduces $N_{\Lambda^3}=\tfrac{1}{2}$ 3-rd rank anti-symmetric. \subref{subfig:branes_SU6_Nf=17}: a stack of 6 D6 branes in between one O8${}^\ast$ (with 9 D8s and two stuck half NS5s) and one O8${}^-$ (with 8 D8s and one stuck half NS5). Thus, there is an $\surm(6)$ vector, $N_f=17$ fundamentals from the D8s, $N_{\Lambda^2}=1$ 2nd-rank anti-symmetric from the NS5-O8${}^-$ boundary, and $N_{\Lambda^3}=\tfrac{1}{2}$ 3-rd rank anti-symmetric from the  NS5-O8${}^\ast$ boundary. \subref{subfig:branes_Sp3_Nf=17-5}: a stack of 6 D6 in between one O8${}^\ast$ (with 9 D8s and two stuck half NS5s) and one O8${}^-$ orientifold (with 8 D8s). As the D6s pass through the O8${}^-$, there is an $\sprm(3)$ vector. The D8s together with the  NS5-O8${}^\ast$ boundary lead to $N_f=17+\tfrac{1}{2}$ vectors and another $N_{\Lambda^3}=\tfrac{1}{2}$ 3-rd rank anti-symmetric.
    }
    \label{fig:branes_SU-Sp_special}
\end{figure}

There are three special SU/Sp theories \eqref{eqn:Special_models} possible on a $0$ curve \cite[Tab.\ 2]{Bhardwaj:2015oru}. For brevity, let us use the following short-hand notation:
\begin{itemize}
    \item $\boldsymbol{\surm(6)^{\prime\prime}}$: $\surm(6)$ gauge theory with $N_f= 18$, $N_{\Lambda^3} =1$
    \item $\boldsymbol{\surm(6)^{\prime}}$: $\surm(6)$ gauge theory with $N_f= 17$, $N_{\Lambda^2} =1 $, $N_{\Lambda^3} =\frac{1}{2}$.
    \item $\boldsymbol{\sprm(3)^{\prime}}$: $\sprm(3)$ gauge theory with $N_f= 17+\frac{1}{2}$, $N_{\Lambda^3} =\frac{1}{2}$.
\end{itemize}
In terms of brane systems, these theories can be realised by a Type I${}^\prime$ construction \cite{Morrison:1996xf,Douglas:1996xp,Gorbatov:2001pw}, see also \cite{Ohmori:2015tka}. Therein, an O8${}^\ast$ is used as one or both of the orientifold planes. We refer to  \cite{Cabrera:2019izd} for a detailed discussion in the magnetic quiver context.  Here, the brane systems are presented in Figure \ref{fig:branes_SU-Sp_special}.

\begin{figure}[ht]\vspace*{-40pt}
\hspace*{-40pt}
    \centering
\includegraphics[page=25,scale=0.9]{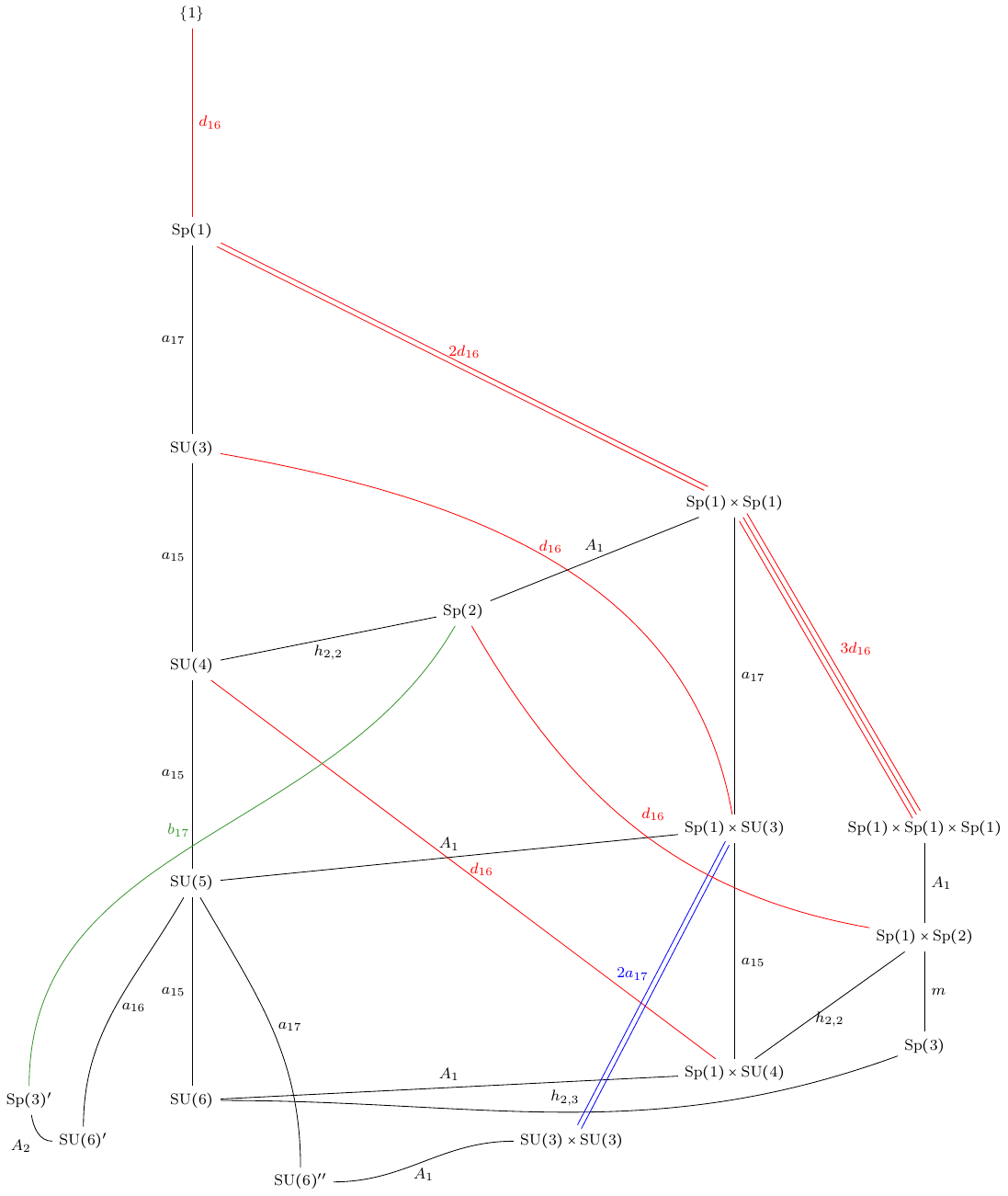}
    \caption{Pattern of Higgs branch RG-flows including special cases theories: $\surm(6)^\prime$ denotes $\surm(6)$ with $N_f= 17$, $N_{\Lambda^2} =1 $, $N_{\Lambda^3} =\frac{1}{2}$. $\surm(6)^{\prime \prime}$ denotes $\surm(6)$ with $N_f= 18$, $N_{\Lambda^3} =1$. $\sprm(3)^{\prime}$ denotes $\sprm(3)$ with $N_f= 17+\frac{1}{2}$, $N_{\Lambda^3} =\frac{1}{2}$. The underlying brane motions  are discussed in Section~\ref{subsec:branes_special}, while the field theory perspective is presented in Section~\ref{subsec:branching_special}.}
    \label{fig:Hasse_with_special}
\end{figure}

As for the general SU/Sp theories, the Higgs branches and RG-flows are analysed below. One observes that these three special cases are connected to the general SU/Sp families via the Higgs branch RG-flows shown in Figure \ref{fig:Hasse_with_special}.

\subsubsection{Partial Higgsing via brane transitions}
\label{subsec:branes_special}
Based on the brane realisations in Figure \ref{fig:branes_SU-Sp_special}, one can analyse the possible brane motions to identify Higgs branch transitions.
\begin{itemize}
    \item \ul{$\surm(6)^{\prime\prime} \to \surm(3) \sqcup \surm(3)$:} The brane system in Figure \ref{subfig:branes_SU6_Nf=18} has two half NS5 branes stuck on each orientifold.  Splitting these NS5 branes along the O-plane (and converting D8 + O8${}^\ast$ $\to$ O8${}^-$), leads to two separated stacks of 3 D6 branes in between stuck half NS5 branes on O8${}^-$ planes. Thus, two copies of the brane system for the standard $\surm(3)$
    theories, see Figure~\ref{subfig:branes_SU}.
    \item \ul{$\surm(6)^{\prime\prime} \to \surm(5)$:} Moving one D6 along the $x^{7,8,9}$ directions off to infinite diminishes the Higgs branch by 17 quaternionic dimensions (one D6 between 18 D8 branes). However, this enforces the conversion D8 + O${8}^\ast$ $\to$ O${8}^-$ in the brane system in Figure \ref{subfig:branes_SU6_Nf=18}, because NS5-O8${}^\ast$ junction does not support an $\surm(5)$ world-volume theory \cite{Gorbatov:2001pw,Zafrir:2015rga,Hayashi:2015zka}. Consequently, one ends up with the  $\surm(5)$ brane system of Figure \ref{subfig:branes_SU}. 
     \item \ul{$\surm(6)^{\prime} \to \sprm(3)^{\prime}$:} The brane system in Figure \ref{subfig:branes_SU6_Nf=17} has one O8${}^-$ plane with an stuck half NS5 branes on it. Moving this half NS5 along the orientifold away from the brane stack off to infinity forces the stack of D6 branes to reconnect and pass through the O8${}^-$. Therefore, the world-volume theory is projected to $\sprm(3)^\prime$ and the brane systems coincide with that of Figure \ref{subfig:branes_Sp3_Nf=17-5}. The motion of the NS5 brane contributes one quaternionic dimension.
     \item \ul{$\sprm(3)^{\prime} \to \sprm(2)$:} Trying to move one D6 along the $x^{7,8,9}$ directions off to infinity requires a second D6, and diminishes the Higgs branch by 32 quaternionic dimensions; this is because the D6 has to be suspended between the 17 D8s, and also respect the boundary conditions on the O8${}^-$. The latter enforces that a second D6 has to be suspended as well. Again, this enforces the conversion D8 + O8${}^\ast$ $\to$ O8${}^-$ in the brane system in Figure \ref{subfig:branes_Sp3_Nf=17-5}, because NS5-O8${}^\ast$ junction does not support an $\sprm(2)$ world-volume theory \cite{Gorbatov:2001pw}. As a result, the $\sprm(2)$ brane system of Figure \ref{subfig:branes_Sp} emerges. 
\end{itemize}
In the subsequent sections, these results are cross-examined by branching rules, quiver subtraction, the decay and fission algorithm, and F-theory geometry. 

\subsubsection{Partial Higgsing via branching rules}
\label{subsec:branching_special}
\begin{itemize}
    \item $\boldsymbol{\surm(6)}$, $\boldsymbol{N_f=18}$, $\boldsymbol{N_{\Lambda^3}=1}$. The following partial Higgs mechanisms are consistent:
\begin{itemize}
    \item \ul{$\surm(6) \to \surm(5)$}:
    \begin{subequations}
    \begin{align}
    [1,0,0,0,0]_{A} &\to   [1,0,0,0]_{A} +  [0,0,0,0]_{A} \\
    [0,0,1,0,0]_{A} &\to  [0,1,0,0]_{A} +  [0,0,1,0]_{A} \\
     [1,0,0,0,1]_{A} &\to     [1,0,0,1]_{A} + [1,0,0,0]_{A} + [0,0,0,1]_{A}  + [0,0,0,0]_{A}  
    \end{align}
    such that the residual $\surm(5)$ has $18-2=16$ fundamentals and $2$ 2nd rank anti-symmetric hypermultiplets. The transition is locally characterised by a $\urm(1)$ gauge group (the commutant of $\surmL(5)$ inside $\surmL(6)$) with $18-1=17$ matter fields, i.e.\ the $a_{17}$ transition.
    \end{subequations}
    \item \ul{$\surm(6) \to \surm(3) \times \surm(3)$}:
    \begin{subequations}
    \begin{align}
    [1,0,0,0,0]_{A} &\to   [1,0]_{A} \otimes [0,0]_{A} +  [0,0]_{A} \otimes [1,0]_{A} \\
    [0,0,1,0,0]_{A} &\to    2 [0,0]_{A} \otimes [0,0]_{A} +  [1,0]_{A} \otimes [0,1]_{A}  +  [0,1]_{A} \otimes [1,0]_{A}\\
     [1,0,0,0,1]_{A} &\to     [1,1]_{A} \otimes [0,0]_{A} + [0,0]_{A} \otimes [1,1]_{A}+ [0,0]_{A} \otimes [0,0]_{A} \\
     &\qquad +  [1,0]_{A} \otimes [0,1]_{A}  +  [0,1]_{A} \otimes [1,0]_{A}   \notag 
    \end{align}
    each of the residual $\surm(3)$ factors has $18$ fundamentals. The commutant of $\surmL(3) \times\surmL(3)$ inside $\surmL(6)$ is $\urmL(1)$, and there are two singlets from the matter fields. Thus, this $1$-dimensional transition is described by $\urm(1)$ with $2$ fundamentals, i.e.\ $A_1 \cong a_1$.
    \end{subequations}
\end{itemize}

\item $\boldsymbol{\surm(6)}$, $\boldsymbol{N_f=17}$, $\boldsymbol{N_{\Lambda^2}=1}$, $\boldsymbol{N_{\Lambda^3}=\frac{1}{2}}$. The following partial Higgs mechanisms are consistent:
\begin{itemize}
    \item \ul{$\surm(6) \to \surm(5)$}:
    \begin{subequations}
    \begin{align}
    [1,0,0,0,0]_{A} &\to   [1,0,0,0]_{A} +  [0,0,0,0]_{A} \\
    [0,1,0,0,0]_{A} &\to  [0,1,0,0]_{A} +  [1,0,0,0]_{A} \\
    [0,0,1,0,0]_{A} &\to  [0,1,0,0]_{A} +  [0,0,1,0]_{A} \\
     [1,0,0,0,1]_{A} &\to     [1,0,0,1]_{A} + [1,0,0,0]_{A} + [0,0,0,1]_{A}   + [0,0,0,0]_{A}  
    \end{align}
    \end{subequations}
    and the residual $\surm(5)$ has $17+1-2=16$ fundamentals and $1 + 2  \cdot \frac{1}{2}=2$ 2nd rank anti-symmetric hypermultiplets. The transition is locally characterised by a $\urm(1)$ gauge group with $17-1=16$ matter fields, i.e.\ the $a_{16}$ transition.
     \item  \ul{$\surm(6) \to \sprm(3)$}:
    \begin{subequations}
    \begin{align}
    [1,0,0,0,0]_{A} &\to   [1,0,0]_{C}  \\
    [0,1,0,0,0]_{A} &\to  [0,0,0]_{C} +  [0,1,0]_{C} \\
    [0,0,1,0,0]_{A} &\to  [1,0,0]_{C} +  [0,0,1]_{C} \\
     [1,0,0,0,1]_{A} &\to     [2,0,0]_{C} + [0,1,0]_{C}
    \end{align}
    \end{subequations}
    such that the residual $\sprm(3)$ theory has $17+\frac{1}{2}$ fundamentals and a $\frac{1}{2}$ 3rd rank anti-symmetric. The transition is of dimension $1$.
\end{itemize}
\item $\boldsymbol{\sprm(3)}$, $\boldsymbol{N_f=17+\frac{1}{2}}$, $\boldsymbol{N_{\Lambda^3}=\frac{1}{2}}$. The following partial Higgs mechanisms are consistent:
\begin{itemize}
    \item \ul{$\sprm(3) \to \sprm(2)$}: First consider an intermediate step $\sprm(3) \to \sprm(1) \times \sprm(2)$:
   \begin{subequations}
    \begin{align}
[1,0,0]_C &\to  [0]_C \otimes[1,0]_C + [1]_C \otimes [0,0]_C\\
[0,0,1]_C &\to  [0]_C \otimes[1,0]_C + [1]_C \otimes [0,1]_C\\
[2,0,0]_C &\to [2]_C \otimes [0,0]_C + [0]_C \otimes[2,0]_C +[1]_C \otimes [1,0]_C
    \end{align}
    \end{subequations}
    which does not give rise to a consistent Higgs mechanism. However, one verifies that further breaking the $\sprm(1)$ \emph{does} lead to a consistent Higgsing  $\sprm(3) \to  \sprm(2)$:
      \begin{subequations}
    \begin{align}
[1,0,0]_C &\to  [1,0]_C + 2 \cdot [0,0]_C\\
[0,0,1]_C &\to  [1,0]_C + 2 \cdot [0,1]_C\\
[2,0,0]_C &\to 3 \cdot [0,0]_C + [2,0]_C +2 \cdot [1,0]_C
    \end{align}
    \end{subequations}
    such that the residual $\sprm(2)$ gauge theory has $(17 + \frac{1}{2}) + \frac{1}{2} -2  = 16$ fundamentals, and $2 \cdot \frac{1}{2}=1 $ 2nd rank anti-symmetrics. The transition is of dimension $2\cdot (17 +\frac{1}{2}) -3 = 32$, and turns out to be a $b_{17}$ transition, see below.
\end{itemize}
\end{itemize}
\FloatBarrier

\subsubsection{Higgs branch Hasse diagram via quiver subtraction}
\label{subsec:QuiverSub_special}
\begin{sidewaysfigure}
    \centering
    \begin{subfigure}[t]{0.485\textwidth}
    \centering
 \includegraphics[page=29,scale=0.75]{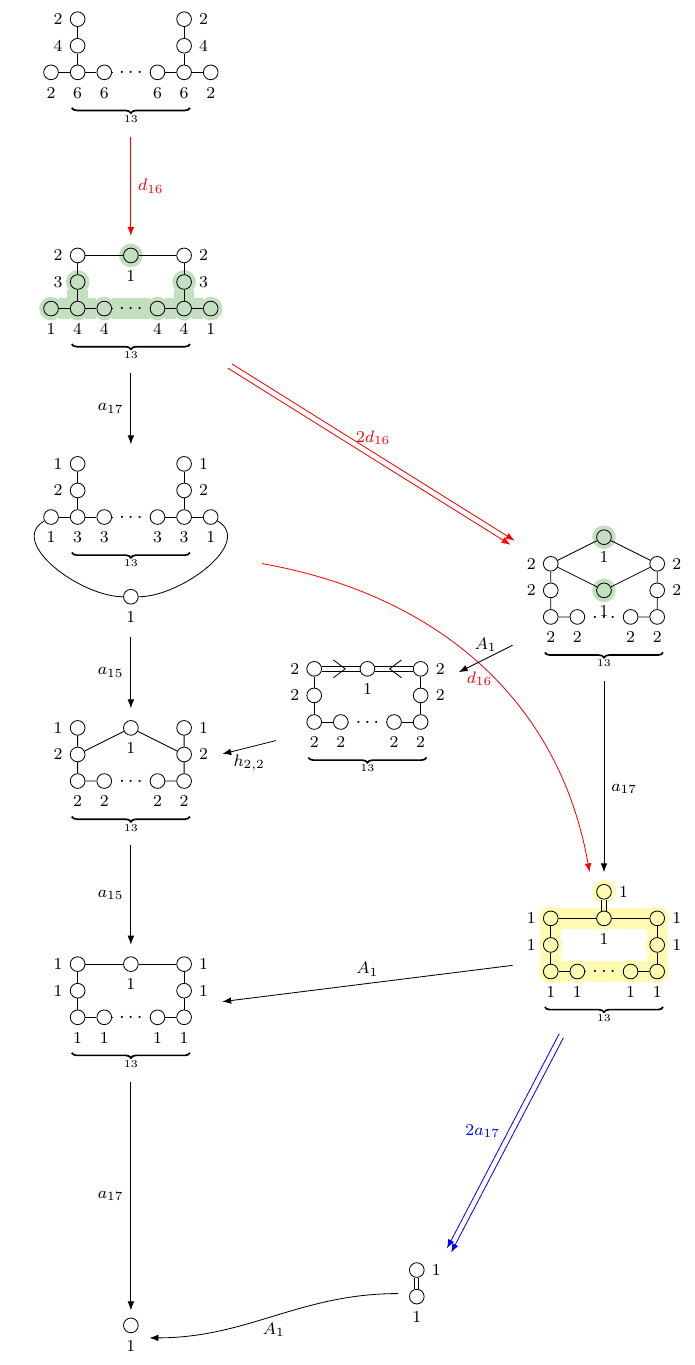}
  \caption{Quiver subtraction}
  \label{subfig:MQ_sub_SU6_18fund}
    \end{subfigure}
    \hfill
     \begin{subfigure}[t]{0.485\textwidth}
    \centering
  \includegraphics[page=20,scale=0.75]{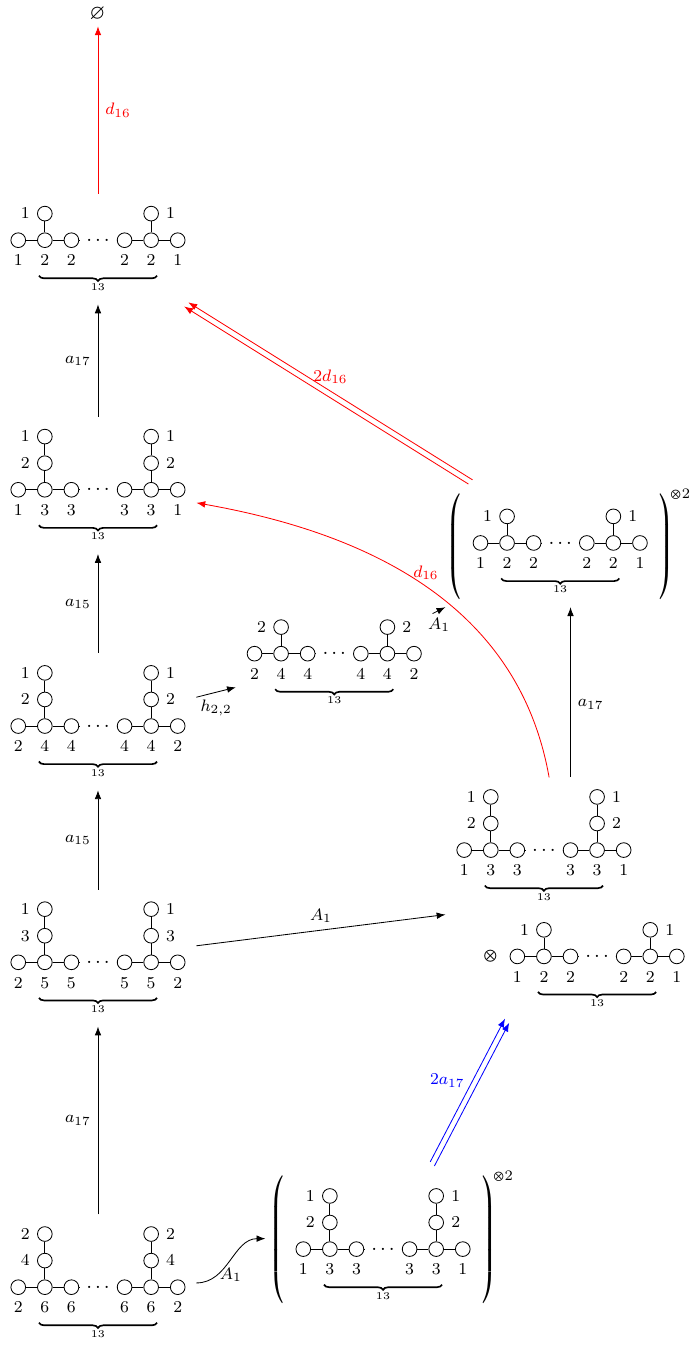}
  \caption{Decay and fission}
  \label{subfig:Decay_fission_SU6_18fund}
    \end{subfigure}
    \caption{Hasse diagram for $\surm(6)$ with $N_f= 18$, $N_{\Lambda^3} =1$. \subref{subfig:MQ_sub_SU6_18fund} via quiver subtraction and \subref{subfig:Decay_fission_SU6_18fund} via the decay and fission algorithm.}
    \label{fig:Hasse_SU6_18fund}
\end{sidewaysfigure}

\begin{sidewaysfigure}
    \centering
    \begin{subfigure}[t]{0.485\textwidth}
    \centering
 \includegraphics[page=30,scale=0.75]{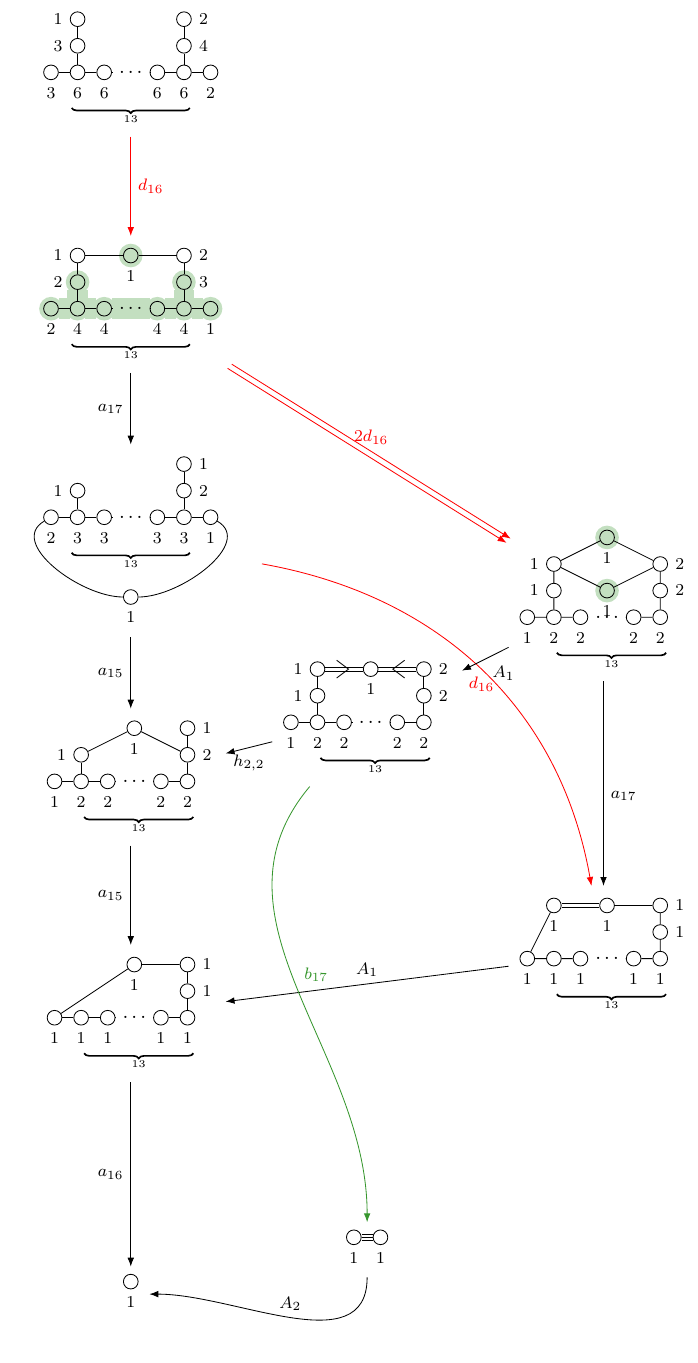}
  \caption{Quiver subtraction}
  \label{subfig:MQ_sub_SU6_17fund}
    \end{subfigure}
    \hfill
     \begin{subfigure}[t]{0.485\textwidth}
    \centering
  \includegraphics[page=21,scale=0.75]{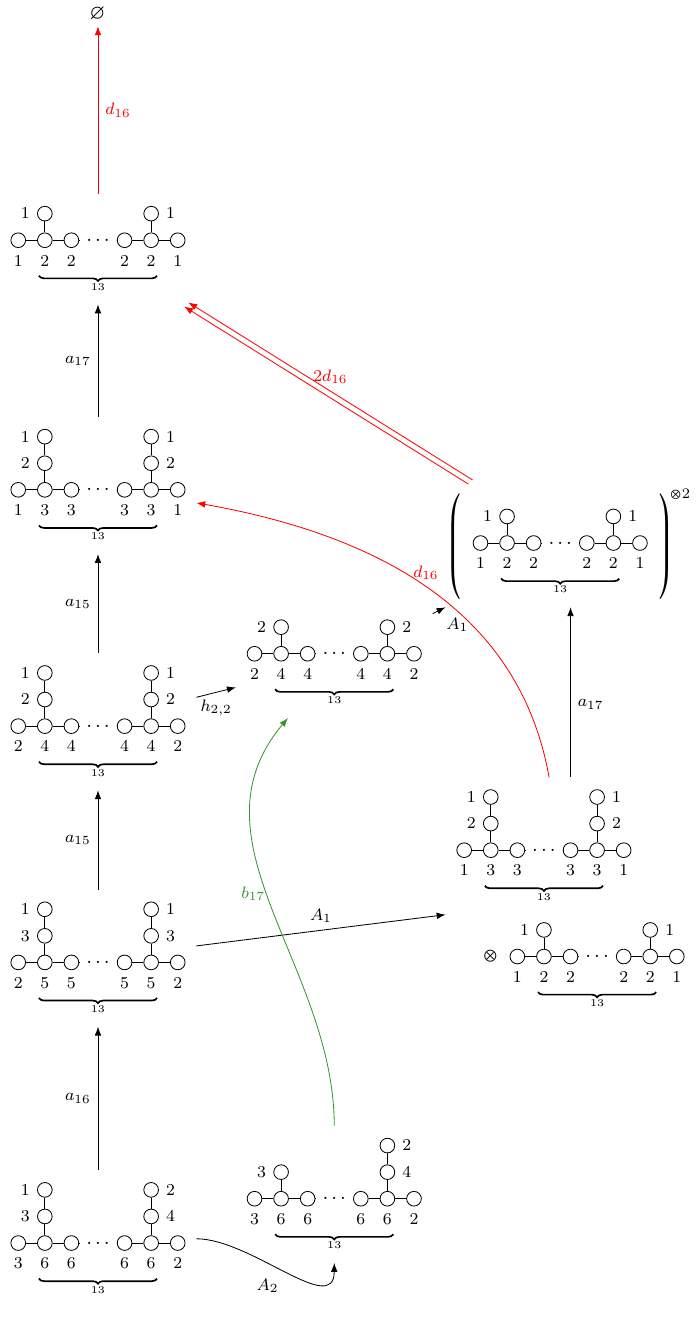}
  \caption{Decay and fission}
  \label{subfig:Decay_fission_SU6_17fund}
    \end{subfigure}
    \caption{Hasse diagram for $\surm(6)$ with $N_f= 17$, $N_{\Lambda^2} =1$, $N_{\Lambda^3} =\frac{1}{2}$.
    \subref{subfig:MQ_sub_SU6_17fund} via quiver subtraction and \subref{subfig:Decay_fission_SU6_17fund} via the decay and fission algorithm.}
    \label{fig:Hasse_SU6_17fund}
\end{sidewaysfigure}

\begin{sidewaysfigure}
    \centering
    \begin{subfigure}[t]{0.485\textwidth}
    \centering
 \includegraphics[page=31,scale=0.75]{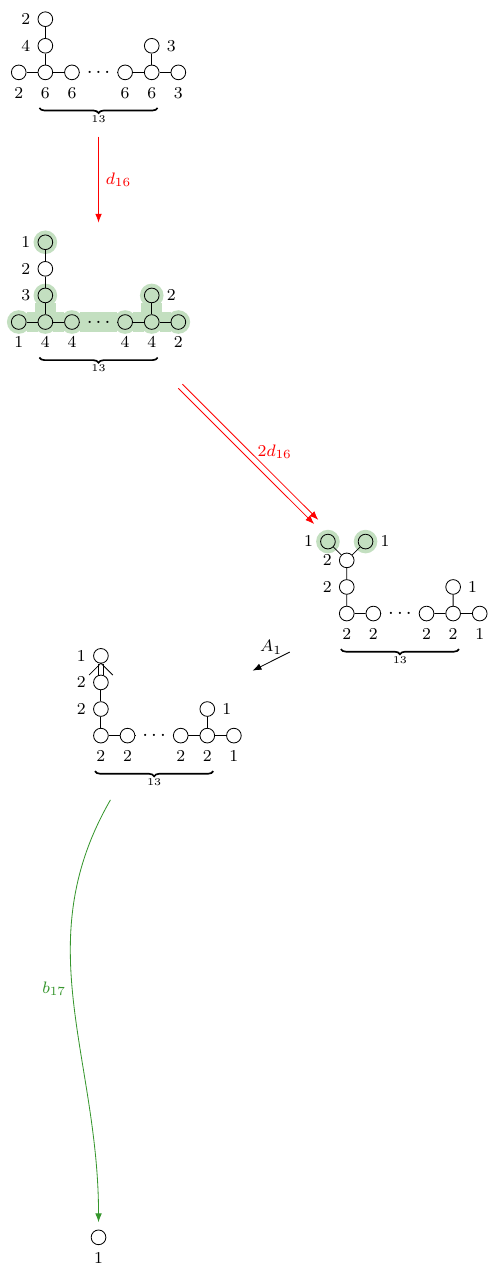}
  \caption{Quiver subtraction}
  \label{subfig:MQ_sub_Sp3_17-half_fund}
    \end{subfigure}
    \hfill
     \begin{subfigure}[t]{0.485\textwidth}
    \centering
  \includegraphics[page=22,scale=0.75]{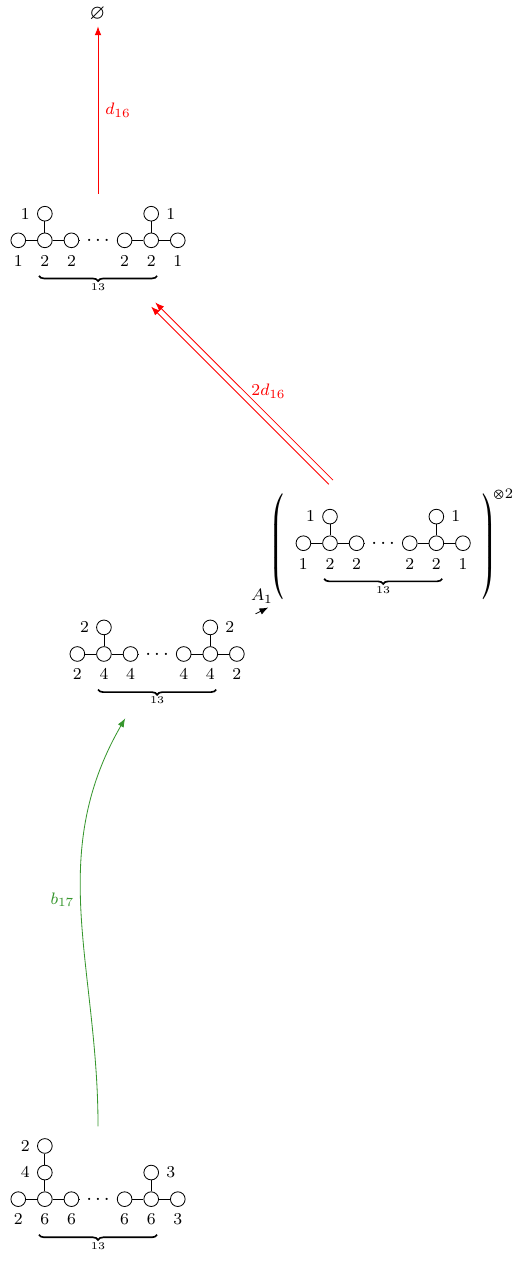}
  \caption{Decay and fission}
  \label{subfig:Decay_fission_Sp3_17-half_fund}
    \end{subfigure}
    \caption{Hasse diagram for $\sprm(3)$ with $N_f= 17+\frac{1}{2}$, $N_{\Lambda^3} =\frac{1}{2}$. \subref{subfig:MQ_sub_Sp3_17-half_fund} via quiver subtraction and \subref{subfig:Decay_fission_Sp3_17-half_fund} via the decay and fission algorithm.}
    \label{fig:Hasse_Sp3_17-half_fund}
\end{sidewaysfigure}

Building on the insights gained in the section above and the rules of \cite{Cabrera:2019izd}, one derives the magnetic quivers for the three special SU/Sp little string theories  from the brane systems in Figure \ref{fig:branes_SU-Sp_special}:
\begin{itemize}
    \item $\surm(6)$ with $N_f= 18$, $N_{\Lambda^3} =1$
    \begin{align}
        \raisebox{-.5\height}{
            \includegraphics[page=26]{pics/0-curve_figures.pdf}
        }
        \label{eq:MQ_SU6_18fund}
    \end{align}
    \item $\surm(6)$ with $N_f= 17$, $N_{\Lambda^2} =1 $, $N_{\Lambda^3} =\frac{1}{2}$
    \begin{align}
        \raisebox{-.5\height}{
            \includegraphics[page=27]{pics/0-curve_figures.pdf}
        }
        \label{eq:MQ_SU6_17fund}
    \end{align}
    \item $\sprm(3)$ with $N_f= 17+\frac{1}{2}$, $N_{\Lambda^3} =\frac{1}{2}$
    \begin{align}
        \raisebox{-.5\height}{
            \includegraphics[page=28]{pics/0-curve_figures.pdf}
        }
         \label{eq:MQ_Sp3_17-half_fund}
    \end{align}
\end{itemize}
Using quiver subtraction, one derives the Higgs branch Hasse diagrams displayed in Figures \ref{subfig:MQ_sub_SU6_18fund}, \ref{subfig:MQ_sub_SU6_17fund}, and \ref{subfig:MQ_sub_Sp3_17-half_fund}. These confirm, on the one hand side, the analysis of partial Higgsing via branching rules and brane motions and, on the other hand, allow to specify the geometry the transitions. The identical Hasse diagrams are recovered independently by the decay and fission algorithm, see Figures \ref{subfig:Decay_fission_SU6_18fund}, \ref{subfig:Decay_fission_SU6_17fund}, and \ref{subfig:Decay_fission_Sp3_17-half_fund} as well as the discussion below.

In addition, the global symmetry can be derived by standard techniques: Appendix \ref{app:monopole} details the monopole operators that furnish the adjoint representation of the global symmetry.

\pagebreak

\FloatBarrier

\subsubsection{Higgs branch Hasse diagram via decay and fission}
Next, the \emph{decay and fission} algorithm \cite{Bourget:2023dkj,InvQuiverSubHiggs} is applied on the magnetic quivers \eqref{eq:MQ_SU6_18fund}--\eqref{eq:MQ_Sp3_17-half_fund} of the special SU/Sp little string theories.
\begin{itemize}
    \item $\boldsymbol{\surm(6)}$, $\boldsymbol{N_f=18}$, $\boldsymbol{N_{\Lambda^3}=1}$. The algorithm outputs Figure \ref{subfig:Decay_fission_SU6_18fund}, which can be detailed as follows:
\begin{compactitem}
    \item \ul{$\surm(6) \to \surm(5)$}: The quiver \eqref{eq:MQ_SU6_18fund} contains a linear chain of $17$ balanced nodes; thus, one finds an $a_{17}$ decay
     \begin{align}
\raisebox{-.5\height}{
 \includegraphics[page=9]{pics/0-curve_IQS_figures.pdf}
 }
\end{align}
and the resulting magnetic quiver is that of the $\surm(5)$ theory, cf.\ \eqref{eq:MQ_SU-odd}.
    \item \ul{$\surm(6) \to \surm(3) \times \surm(3)$}: The quiver \eqref{eq:MQ_SU6_18fund} also contains a stack of 2 affine $d_{16}$ Dynkin diagrams; however, since this stack is connected to balanced gauge nodes, one cannot fission it this way. Instead, observe that \eqref{eq:MQ_SU6_18fund} can fission into exactly two copies of the magnetic quiver for the $\surm(3)$ theory \eqref{eq:MQ_SU-odd}. Therefore, the fission  reads 
     \begin{align}
\raisebox{-.5\height}{
 \includegraphics[page=10]{pics/0-curve_IQS_figures.pdf}
 }
\end{align}
and the resulting magnetic quivers are those of the $\surm(3)$ theory. The transition geometry is detailed in \cite{InvQuiverSubHiggs}.
\end{compactitem}

\item $\boldsymbol{\surm(6)}$, $\boldsymbol{N_f=17}$, $\boldsymbol{N_{\Lambda^2}=1}$, $\boldsymbol{N_{\Lambda^3}=\frac{1}{2}}$. The algorithm generates Figure \ref{subfig:Decay_fission_SU6_17fund}; in more detail:
\begin{compactitem}
    \item \ul{$\surm(6) \to \surm(5)$}: The magnetic quiver \eqref{eq:MQ_SU6_17fund} contains a linear chain of $16$ balanced gauge nodes. Therefore, the decay becomes
     \begin{align}
\raisebox{-.5\height}{
 \includegraphics[page=11]{pics/0-curve_IQS_figures.pdf}
 }
\end{align}
such that the resulting magnetic quiver is that of the $\surm(5)$ theory \eqref{eq:MQ_SU-odd}.
     \item  \ul{$\surm(6) \to \sprm(3)$}: The quiver \eqref{eq:MQ_SU6_17fund} also contains an overbalanced $\urm(1)$ node, with balance $b=1$, connected to another node with positive balance. This leads to a $A_{b+1}$ decay \cite{Bourget:2023dkj,InvQuiverSubHiggs}
     \begin{align}
\raisebox{-.5\height}{
 \includegraphics[page=12]{pics/0-curve_IQS_figures.pdf}
 }
\end{align}
such that the resulting magnetic quiver is that of the special $\sprm(3)$ theory \eqref{eq:MQ_Sp3_17-half_fund}.
\end{compactitem}
\item $\boldsymbol{\sprm(3)}$, $\boldsymbol{N_f=17+\frac{1}{2}}$, $\boldsymbol{N_{\Lambda^3}=\frac{1}{2}}$. The algorithm produces Figure \ref{subfig:Decay_fission_SU6_18fund}, which follows from:
\begin{compactitem}
    \item \ul{$\sprm(3) \to \sprm(2)$}: The quiver \eqref{eq:MQ_Sp3_17-half_fund} 
decays as follows:
\begin{align}
\raisebox{-.5\height}{
 \includegraphics[page=14]{pics/0-curve_IQS_figures.pdf}
 }
\end{align}
such that the resulting magnetic quiver is that of the $\sprm(2)$ theory \eqref{eq:MQ_Spk}. The transition is identified as $b_{17}$ \cite{InvQuiverSubHiggs}.
\end{compactitem}
\end{itemize}
Again, the \emph{decay and fission} algorithm is in agreement with the Higgs branch RG-flows obtained via brane motions in Section~\ref{subsec:branes_special} and branching rules in Section~\ref{subsec:branching_special}. Moreover, the algorithm also confirms the geometry of the transverse slices obtained via quiver subtraction in Section~\ref{subsec:QuiverSub_special}.

\section{F-Theory perspective on the Hasse diagram and T-duality}
\label{sec:F-theory}

\subsection{Charting the Hasse diagram via geometry}
The technique of magnetic quiver reveals itself as a powerful tool to extract the full Hasse diagram of a six-dimensional theory with minimal supersymmetry 
\cite{Cabrera:2019izd,Cabrera:2019dob,Sperling:2021fcf,Hanany:2022itc,Fazzi:2022hal,Bourget:2022tmw,Fazzi:2022yca}. Nevertheless, one has to be careful when applying it as it may be non-trivial to assign an electric theory to each leaf in the foliation/stratification. In more detail, (standard) quiver subtraction provides magnetic quivers for the leaves in the foliation; such that these quivers are \emph{not} the magnetic quivers of the theories the RG-flows end on. The decay and fission algorithm \cite{Bourget:2023dkj,InvQuiverSubHiggs} resolves the short-coming as the generated quivers \emph{are} the magnetic quivers of the RG-flow end points. 
Nonetheless, it is still a formidable task to identify which electric theory a given magnetic quiver corresponds to; especially, if the former is not (yet) fully detailed.
That is to say, for each Higgs branch RG-flow deduced via magnetic quivers, one must always be able to derive an F-theory description.

In fact, we recall that the F-theory description requires the specification of a fibration over a base space $\mathcal{B}$ that needs to be compactified in order for the six-dimensional theory to emerge. In this language, running a Higgs branch RG-flow translates to turning on complex structure deformations whose action usually consists in a smoothing of the singularity type, even if there may be geometric deformations with a non-trivial physical realisation \cite{Heckman:2015ola,Alvarez-Garcia:2023gdd}.

\begin{sidewaysfigure}
    \centering
    \includegraphics[page=17]{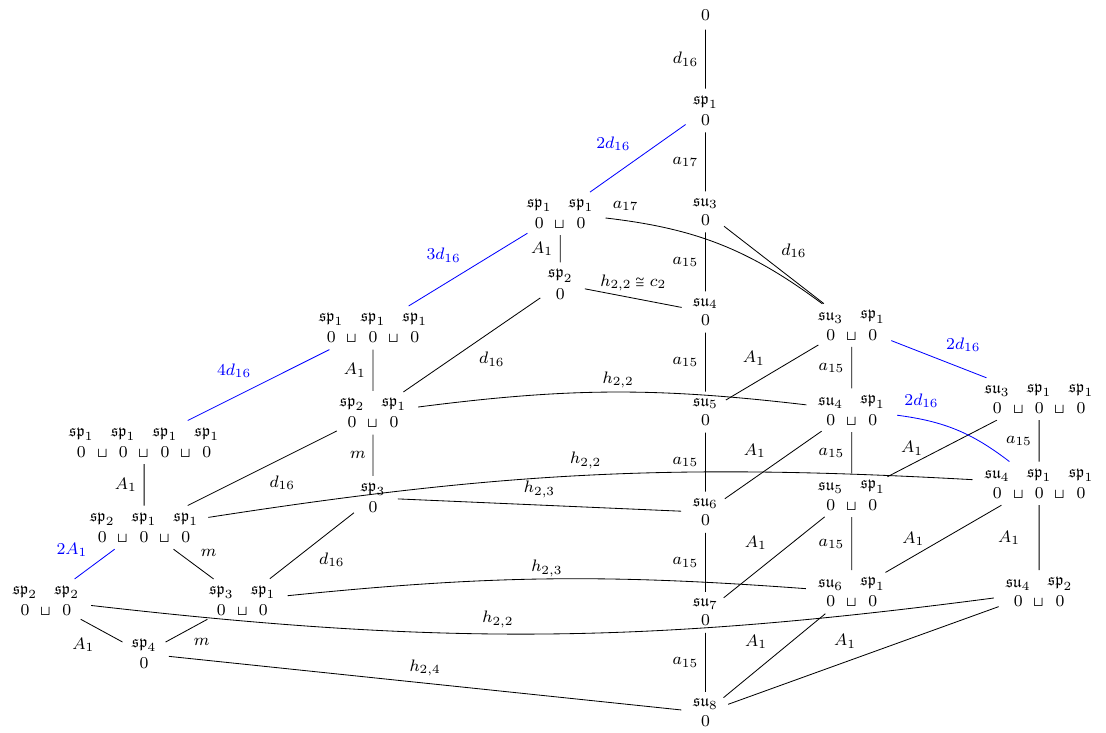}
    \caption{Hasse diagram for the {$\footnotesize \stackon{$0$}{$\mathfrak{su}_8$}$} LST obtained from complex structure deformation of the geometrical engineering. The name, and the multiplicity of blue slices, of the various slices have been assigned thanks to the Hasse diagram extracted from the Magnetic Quiver via Quiver subtraction.}
    \label{fig:Hasse_0su8_Ftheory}
\end{sidewaysfigure}

As pointed out in Section~\ref{sec:Brief_LST}, the zero curve cases studied in this paper are realised as a fibration over a $\mathbb{P}^1$ space with normal bundle $\left(T^2\times \mathbb{C}\right)$ decorated with an $I_N$ fibre that can be either \emph{split} $I_{N}^s$, realising an $\mathfrak{su}_N$ gauge algebra, or \emph{non-split} $I^{ns}_N$, realising an $\mathfrak{sp}_{\lfloor \frac{N}{2} \rfloor}$ gauge algebra \cite{Aspinwall:2000kf,Bhardwaj:2015oru}. In order to understand how to extract the full Hasse diagram for the Higgs branch of this theory directly from geometry, we can break down the various branchings met in Section~\ref{sec:Branes-MQ} and re-map them to the respective intuitive geometric realisation: 
\begin{itemize}
    \item \ul{$\surm(2N) \rightarrow \sprm(N)$}. Geometrically this process has the easiest realisation among those commented on here. In fact, it simply is a change of monodromy for the $I_{2N}$ fibre \cite{Bershadsky:1996nh}. Explicitly, consider the generic Weierstrass model:
    \begin{equation}\label{eqn:generic_Weierstrass}
        y^2+a_1 xy +a_3 y = x^3 + a_2 x^2 + a_4 x +a_6 \ .
    \end{equation}
    The coefficients $a_i$ are polynomial in local coordinates of the base space and their degree, together with the one of the discriminant $\Delta$ of \eqref{eqn:generic_Weierstrass} dictates the fibre structure:
    \begin{equation}
    \begin{tabular}{c|c|c|c|c|c|c}
         & $a_1$ & $a_2$ & $a_3$ & $a_4$  & $a_6$ & $\Delta$  \\ \midrule
       $I_{2N}^{ns}$  & $0$ & $0$ & $N$ & $N$ & $2N$ & $2N$ \\
       $I_{2N}^{s}$  & $0$ & $1$ & $N$ & $N$ & $2N$ & $2N$ \\
    \end{tabular}    
    \end{equation}
    Thus the splitness of the fibre is regulated by the existence of a factorisation of the polynomial
    \begin{equation}\label{eqn:PolynomialSplitness}
    y^2+a_1 xy-a_2x^2    \ ,
    \end{equation}
    which can be affected by a suitable complex structure deformation.
    
    \item \ul{$\surm(N+1) \rightarrow \surm(N)$}. This process is characterised by the Higgsing of the flavour symmetry originating from the matter in the fundamental representation of the gauge algebra. Thus, we expect it to realise a simple smoothing of the fibre's singularity. In fact, let us first consider the vanishing order of the coefficients of \eqref{eqn:generic_Weierstrass} for an $I^{s}_{2k+1}$ split fibre:
        \begin{equation}
        \begin{gathered}
            \begin{tabular}{c|c|c|c|c|c|c}
         & $a_1$ & $a_2$ & $a_3$ & $a_4$  & $a_6$ & $\Delta$  \\ \midrule
       $I_{2k}^{s}$  & $0$ & $1$ & $k$ & $k$ & $2k$ & $2k$ \\
       $I_{2k+1}^{s}$  & $0$ & $1$ & $k$ & $k+1$ & $2k+1$ & $2k+1$ \\
    \end{tabular}  
        \end{gathered} 
    \end{equation}
    It is clear that starting from $N=2k$, the Higgsing $\surm(N+1) \rightarrow \surm(N)$ stems from a complex structure deformation that smoothens the singular order of $a_4$, $a_6$, and $\Delta$. An analogous discussion holds for $N=2k-1$, as the deformation reduces the singularity of the coefficients $a_3$, $a_6$, and $\Delta$.

    \item \ul{$\surm(2N+M)\rightarrow \sprm(N) \, \sqcup \, \surm(M)$} \emph{and} \ul{$\sprm(N+M)\rightarrow \sprm(N) \, \sqcup \, \sprm(M)$}. In the Type IIA realisation both of these Higgsings account to separating a stack of $N+M$ coincident D6 branes into two subsets of $N$ and $M$ coincident branes and taking the infinite distance limit. A similar example of infinite-distance splitting has already been found in the study of Orbi-Instanton theories\footnote{The $(\mathfrak{e}_8,\mathfrak{g}_{ADE})$ Orbi-Instanton theory is the 6d $\mathcal{N}=1$ SCFT living on the world-volume of coincident M5 branes probing an M9 end-of-the-world wall and a $\mathbb{C}^2/\Gamma_{ADE}$ orbifold singularity, where $\Gamma_{ADE}\subset \surm(2)$ is a finite subgroup of $\surm(2)$ related to $\mathfrak{g}_{ADE}$ by McKay correspondence. When the orbifold is trivial the theory is also known as \emph{higher rank E-string}, and there exist Higgs branch RG-flows corresponding to the separation, along the M9 plane direction, of a stack of M5 branes in two different stacks \cite{Lawrie:2023uiu}.} in \cite{Heckman:2015ola, Lawrie:2023uiu}, and in both cases the decay and fission algorithm \cite{Bourget:2023dkj} correctly reproduce the results of this separation-action in a one-step process.
    
    The geometrical counterpart of this splitting is not straightforward, as it comprises an intermediate step and a tuning step to realise physical models:
    \begin{itemize}
        \item \emph{The intermediate }\ul{$\surm(N+M)\rightarrow \surm(N) \, \sqcup \, \surm(M)$} \emph{step}.  This transitory move has been studied in \cite{Katz:1996xe}, and is based on the possible resolutions of an $A_{N+M-1}$ singularity:
    \begin{equation}\label{eqn:A_{n+m-1}inA_{n-1}cupA_{m}}
        xy+ \sum \limits_{i=1}^{N+M} (z-t_i)=0 \ ,
    \end{equation}
    where $t_i$ are deformation parameters that we can tune. Therefore, choosing $N$ of the parameters equal to $t$, and the remaining $M$ to be trivial, realises the singular space:
    \begin{equation}\label{eqn:A_{n+m-1}inA_{n-1}cupA_{m-1}}
        xy+ z^{M} \left( z-t \right)^N =0 \ .
    \end{equation}
    The space in \eqref{eqn:A_{n+m-1}inA_{n-1}cupA_{m-1}} has two singularities: an $A_{M-1}$ singularity at $(0,0,0)$, and an $A_{N-1}$ singularity in $(0,0,t)$. The fact that the singularities sit in different loci is a clear signal of a cup theory, as required from this engineering.
    
    But at this step, one should ask whether this deformation allows for split fibers in both the singular points. In fact, it can be shown that one of the two loci is always enforced to support a non-split fiber. Because the vanishing order of the $a_2$ term in \eqref{eqn:generic_Weierstrass} is one, the split-like behaviour stays localised in only one point, this can be interpreted brane-wise with the presence of only a single pair of half NS5s in the model.

    \item \emph{Tuning to} \ul{$\surm(2N+M)\rightarrow \sprm(N) \, \sqcup \, \surm(M)$}. Recalling that in the Type IIA realisation, we can think of the $0$-curve theory equipped with an $\sprm(N)$ gauge algebra as a model free of NS5 branes, from a geometry point of view, we are combining the deformation of the singularity $A_{2N+M-1}$ with a suitable local change of coordinates for the newly realised less singular $A_{2N-1}$ space. This renders the polynomial in \eqref{eqn:PolynomialSplitness} explicitly non-factorisable, whilst keeping the polynomial singular in the $A_{M-1}$ locus factorisable.
    
    \item \emph{Tuning to} \ul{$\sprm(N+M)\rightarrow \sprm(N) \, \sqcup \, \sprm(M)$}. Yet again, we are deforming the initial $A_{N+M-1}$ singularity, for which the polynomial in \eqref{eqn:PolynomialSplitness} was non-factorisable by construction, into an $A_{M-1}$ and $A_{N-1}$ singularities. For both singular loci the splitness-regulating polynomial is still non-factorisable: changing the fibres' monodromy would require a non-trivial $a_2$ term for \eqref{eqn:generic_Weierstrass} that, in this case, cannot be introduced via complex structure deformations.
    
    \end{itemize}

\end{itemize}
The geometric approach automatically outputs physical theories; therefore, it should be morally regarded as a \emph{decay and fission} algorithm applied to engineer Higgs branch RG-flows \cite{Bourget:2023dkj,InvQuiverSubHiggs}. In fact, apart from the vanilla moment map Higgsing of the flavour symmetry, which is usually associated with non-product theories, one must always take into account the possibility of realising non-trivial deformations that lead to product theories whenever they may be possible --- much like the \emph{decay} and \emph{fission} processes a magnetic quiver can undergo. As consequence, we can infer the full Hasse diagram of a six-dimensional theory from geometry by keeping in mind these possibilities. For instance, in Figure \ref{fig:Hasse_0su8_Ftheory} we derived the Hasse diagram for a $0$-curve model equipped with an $\mathfrak{su}_8$ gauge algebra from the geometry, associating to each vertex in the Hasse the respective electric quiver directly readable from the F-theory model.
  
\subsection{\texorpdfstring{F-theory Higgsing of theories with $\Lambda^3$ matter}{F-theory Higgsing of theories with 3rd rank anti-symmetric matter}}
As introduced in the Section~\ref{sec:Brief_LST}, there are three zero curves model that can be deemed as special because anomaly cancellation allows matter in the $3^{\mathrm{rd}}$-$\mathrm{rank}$ anti-symmetric representation of $\surm(6)$ or $\sprm(3)$. A natural task would be to encode this different matter assignment and motivate the Higgs branch RG-flows shown in Figure \ref{fig:Hasse_with_special} also from the geometry. In fact, the analysis of the previous section \emph{fails} to encompass the following cases:
    \begin{alignat}{3}
             &[N_{f}= 17]\stackunder{\stackon{$0$}{$\mathfrak{su}_6$}}{$[N_{\Lambda^2}=1]$}[N_{\Lambda^3}= \tfrac{1}{2}] && \ \rightarrow \ &&  \ \ [N_{f}= 17\tfrac{1}{2}]\stackon{$0$}{$\mathfrak{sp}_3$}[N_{\Lambda^3}= \tfrac{1}{2}]  \,, \\
           & \ \ \ \ \, [N_{f}= 18]\stackon{$0$}{$\mathfrak{su}_6$}[N_{\Lambda^3}= 1]  && \ \rightarrow \ && \ \ \ \ \  \stackunder{\stackon{$0$}{$\mathfrak{su}_3$}}{$[N_{f}= 18]$} \sqcup \stackunder{\stackon{$0$}{$\mathfrak{su}_3$}}{$[N_{f}= 18]$} \,, \\
            & \ \ \,  [N_{f}= 17 +\tfrac{1}{2}]\stackon{$0$}{$\mathfrak{sp}_3$}[N_{\Lambda^3}= \tfrac{1}{2}] && \ \rightarrow \ && \ \ \ \ \, [N_{f}= 16]\stackon{$0$}{$\mathfrak{sp}_2$}[N_{\Lambda^2}= 1] \,.
    \end{alignat}
This discussion gives us space to spend a few words on how to associate vanishing order data with Higgs branch RG-flows and matter content. From this, the description of these aforementioned branching directions follows automatically.

In the seminal work \cite{Bershadsky:1996nh}, via an extension of Tate's algorithm \cite{10.1007/BFb0097582}, a non-perturbative enhancement to the gauge symmetries obtained in Heteretotic models have been clarified. Considering the physics of a $\mathbb{P}_\mathrm{fiber}^1$ fibration over $\mathbb{P}^1_\mathrm{base}$ specified by an integer $n$, i.e.\ the Hirzebruch surface $\mathbb{F}_n$, everything --- from the gauge algebra to the matter content and the possible Higgsing directions --- is encoded in the Weierstrass model associated with the fibration.

The starting point is a simplified version of \eqref{eqn:generic_Weierstrass}, obtained by completing the square in $y$ and the cube in the $x$ variable:
\begin{equation}
    y^2=x^3+f(z_1,z_2)x+ g(z_1,z_2) \ \mathrm{ with } \ \Delta=4f^3+27g^2 \,,
\end{equation}
the variable $z_1$ is the coordinate along the $\mathbb{P}_\mathrm{fiber}^1$ fiber, whereas $z_2$ is a coordinate for the $\mathbb{P}^1_\mathrm{base}$ base. Crucially, the polynomial functions $f(z_1,z_2)$ and $g(z_1,z_2)$ can be decomposed according to their vanishing order in the fibration variable:
\begin{subequations}
    \begin{align}
    f(z_1,z_2) &=\sum \limits_{i=0}^I z_1^i f_{8+n(4-i)}(z_2) \,, \\
    g(z_1,z_2) &=\sum \limits_{i=0}^J z_1^i g_{12+n(6-i)}(z_2) \,.
\end{align}
\end{subequations}
The positive integers $I\le8$ and $J\le12$ are determined as the largest integers such that the subscript of the polynomial functions $f$ and $g$, which stands for the highest degree of $z_2$ appearing with non-trivial coefficient therein, is non-negative. In fact, apart from extracting again from the N\`eron-Koidara classification of singular fibres --- hence from the vanishing order of the triple $(f,g,\Delta)$ --- the gauge algebra associated with certain minimal singularities, the matter content is now encoded in the zeroes of the polynomial factors $f_{\cdot}(z_2)$ and $g_{\cdot}(z_2)$ in the discriminant $\Delta$. Thus, the problem of whether different representations of the gauge algebra can appear in the matter content is translated into factorisation properties of these irreducible components.

Now specialising to the $0$-curve case corresponds to taking $n=0$ in the subscripts of the $z_1$-graded polynomial expansion of $f$ and $g$, but we delay the substitution until the very last step in order not to coincidentally create ambiguous polynomial relations. In this setting, the standard $\surm(6)$ theory with $2+n$ hypermultiplets in the $[0,1,0,0,0]_{A_5}$ and $16+2n$ hypermultiplets in the $[1,0,0,0,0]_{A_5}$ is engineered via the polynomials
\begin{equation}
    f_{4+2n}=h^2_{2+n} \quad , \quad f_{8+n} \quad , \quad f_4 \quad , \quad g_{12+3n}=s_{n+4}^3 \,.
\end{equation}
The discriminant locus factorises as:
\begin{equation}
    \Delta={z_1}^6 h^4_{2+n} P_{2n+16} \,,
\end{equation}
such that the zeroes of $h_{2+n}$ correctly reproduce the hypermultiplets in the anti-symmetric representation and the zeroes of $P_{2n+16}$ the number of fundamental ones. The special theories with $r$ multiplets in the $[0,0,1,0,0]_{A_5}$ can be obtained imposing the extra constraint:
\begin{equation}\label{eqn:constraintSU6}
    h_{n+2}= t_r \Tilde{h}_{2+n-r} \,,
\end{equation}
which subdivides the original $n+2$ multiplets in the $[0,1,0,0,0]_{A_5}$ in $n+2-r$ multiples, transforming in the same representation, and $r$ half hypermultiplets transforming in the $[0,0,1,0,0]_{A_5}$.\\
At this stage recovering the special $\sprm(3)$ theory corresponds to imposing non-splitness on the $\surm(6)$ theory with $r$ $[0,0,1,0,0]_{A_5}$ matter, and this accounts to have:
\begin{equation}\label{eqn:constraintSp3}
    f_{4+2n}= t^2_r \Tilde{h}^2_{2n+4-2r} \,.
\end{equation}

Therefore, we explained the Higgsing $\surm(6)'\rightarrow \sprm(3)'$ as a tuning of the $f_{4}$ term of the unhiggsed $\surm(6)'$ special theory with $1$ multiplets in the $[0,0,1,0,0]_{A_5}$ via complex structure deformation. And now, thanks to the further constraint of \eqref{eqn:constraintSU6}, the Higgsing $\surm(6)'' \rightarrow \surm(3) \sqcup \surm(3)$ is understood, and with \eqref{eqn:constraintSp3} the Higgsing $\sprm(3)'\rightarrow \sprm(2)$ is clarified as well.

For details on when and how to impose the aforementioned polynomial relations, we refer the reader to the original work of \cite{Aspinwall:1996vc}, of which we retained both the notation and the conceptual structure.

\subsection{T-Dualities}
As mentioned in Section~\ref{sec:Introduction}, an important property of LSTs is that they exhibit T-duality: compactifying different models along an $S^1$ with certain choices of Wilson lines produces the same five-dimensional theory. This implies that the Higgs branch structure is modified by the symmetry breaking induced by the extended objects upon compactification. Nevertheless, studying the Higgs branch of the six-dimensional model allows us to extract some information about the five-dimensional theory. In \cite{Lawrie:2023uiu} a first step in this direction has been taken: by running a Higgs branch RG-flow between a pair of T-dual models, a full chain of putative T-dual theories can be extracted from the Hasse diagram.

Before venturing into a detailed analysis, it is important to recall that a special class of LSTs, which also includes the zero curve models discussed in this paper, are the Heterotic LSTs: they intrinsically guide us toward identifying T-dual pairs. These theories can be thought of as compactifications of Type~I $E_8 \times E_8$ or $\mathrm{Spin}(32)/\mathbb{Z}_2$ models that are relate to each by T-duality, because of the uniqueness of the self-dual bosonic weight lattice \cite{Narain:1985jj}. The nomenclature for the $E_8 \times E_8$ little string theories has been established in \cite{Lawrie:2023uiu}, the theory
\begin{equation}
    \mathcal{K}_{N_L,N_R,\mathfrak{g}_{ADE}}(\rho_L,\rho_R)
\end{equation}
refers to the LST whose M-theory uplift is realised by $2$ M9 planes within which spans a $\mathbb{C}^2/\Gamma_{ADE}$ singular space, where $\Gamma_{ADE}\subset \surm(2)$ is the finite group of $\surm(2)$ related to $\mathfrak{g}_{ADE}$ by McKay correspondence. For $\mathfrak{g}=\mathfrak{su}_k$ the short-end notation $\mathcal{K}_{N_L,N_R,k}(\rho_L,\rho_R)$ is adopted. The orbifold boundary conditions $\rho_L,\rho_R$ determine an embedding of $\Gamma_{ADE}\rightarrow E_8$ in each of the M9 planes, whereas $N_L$ and $N_R$ correspond to the number of instantonic M5 branes probing the respective end-of-the-world branes and the orbifold.

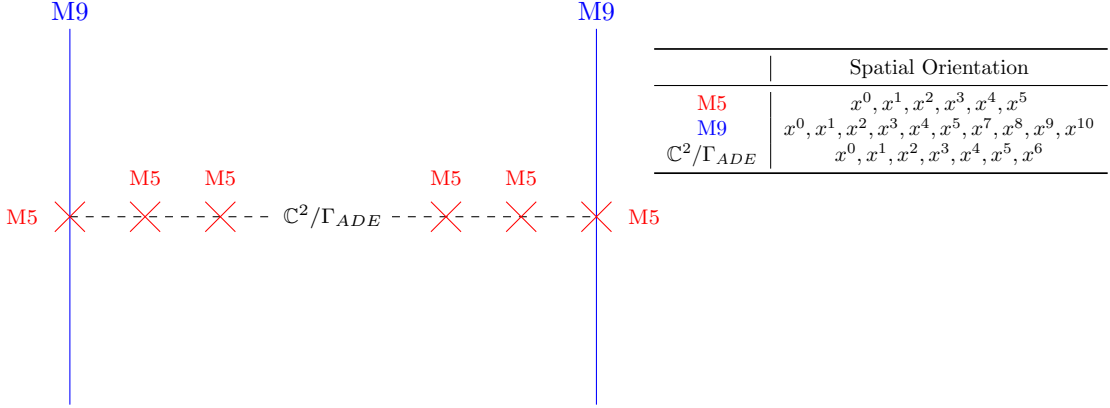
\begin{figure}[ht]
\begin{multicols}{2}
\centering
    \begin{tikzpicture}
    \node[] at (3.5,0) (label) {\footnotesize $\mathbb{C}^2/\Gamma_{ADE}$};
    \draw[dashed] (0,0)--(label)--(7,0);
    \draw[blue] (7,-2.5)--(7,2.5) node[above] () {M9};
    \draw[blue] (0,-2.5)--(0,2.5) node[above] () {M9};
    \node at (7,0) [red,right=0.3] () {\footnotesize M5};
    \draw[red] (6.8,0.2)--(7.2,-0.2);
    \draw[red] (6.8,-0.2)--(7.2,0.2);
    \node at (6,0) [red,above=0.3] () {\footnotesize M5};
    \draw[red] (5.8,0.2)--(6.2,-0.2);
    \draw[red] (5.8,-0.2)--(6.2,0.2);
    \node at (5,0) [red,above=0.3] () {\footnotesize M5};
    \draw[red] (4.8,0.2)--(5.2,-0.2);
    \draw[red] (4.8,-0.2)--(5.2,0.2);
    \node at (2,0) [red,above=0.3] () {\footnotesize M5};
    \draw[red] (1.8,0.2)--(2.2,-0.2);
    \draw[red] (1.8,-0.2)--(2.2,0.2);
    \node at (1,0) [red,above=0.3] () {\footnotesize M5};
    \draw[red] (0.8,0.2)--(1.2,-0.2);
    \draw[red] (0.8,-0.2)--(1.2,0.2);
    \node at (0,0) [red,left=0.3] () {\footnotesize M5};
    \draw[red] (-0.2,0.2)--(0.2,-0.2);
    \draw[red] (-0.2,-0.2)--(0.2,0.2);

    \end{tikzpicture}
    
    \vfill
    
    \begin{equation*}
        \begin{aligned}
        \resizebox{0.4\textwidth}{!}{
                \begin{tabular}{c|c}\toprule
              & Spatial Orientation  \\ \midrule
             {\color{red}M5} & $x^0,x^1,x^2,x^3,x^4,x^5$ \\ 
             {\color{blue}M9} & $x^0,x^1,x^2,x^3,x^4,x^5,x^7,x^8,x^9,x^{10}$ \\ 
             $\mathbb{C}^2/\Gamma_{ADE}$ & $x^0,x^1,x^2,x^3,x^4,x^5,x^6$ \\ \bottomrule
    \end{tabular} 
    }
        \end{aligned}
    \end{equation*}

\end{multicols}
    \caption{The M-theory configuration engineering a $\mathcal{K}_{N_L,N_R,\mathfrak{g}_{ADE}}(\rho_L,\rho_R)$ LST, at a generic point in the tensor branch the M5-branes probe the $\mathbb{C}^2/\Gamma_{ADE}$ orbifold singularity put between two end-of-the-world M9-branes. The singularity is localised in the $x^7,x^8,x^9,x^{10}$ coordinates and spans the remaining ones. }
    \label{fig:M-theory_LST_Heterotic}
\end{figure}

In Figure \ref{fig:M-theory_LST_Heterotic}, the $E_8 \times E_8$ theory is depicted at a generic point of its tensor branch, where the M5-branes are free to probe any point of the orbifold. The reader is referred to \cite{DelZotto:2022ohj,Lawrie:2023uiu} for more details on these models, including their curve configuration, magnetic quiver, and structure constants.

In the T-dual $\mathrm{Spin}(32)/\mathbb{Z}_2$ frame, the central theory \cite{Witten:1995gx} is realised by $k$ Type~I instantonic branes and it is exactly the 
\begin{equation}
    [N_{f}= 16]\stackon{$0$}{$\mathfrak{sp}_N$}[N_{\Lambda^2}= 1]
\end{equation}
little string theory, and all the other $\mathrm{Spin}(32)/\mathbb{Z}_2$ LSTs can be obtained orbifolding this parent model \cite{Blum:1997fw,Blum:1997mm,Intriligator:1997dh}. We refer to \cite{DelZotto:2022ohj,Lawrie:2023uiu} for details on the interplay of the orbifolded $\mathrm{Spin}(32)/\mathbb{Z}_2$ little string theories and the $E_8 \times E_8$ counterpart under T-duality. 

This brief introduction to Heterotic LSTs allows us to perform a T-duality analysis along the lines of \cite{DelZotto:2020sop,DelZotto:2022xrh}, so that we can obtain candidate T-duals to the $0$-curve models, considered in this paper, by searching for theories with matching structure constants, 5d Coulomb branch dimension, and flavour symmetry rank. The starting point is the natural map:
\begin{equation}\label{eqn:Starting_T-duality}
    \mathcal{K}_{N_L,N_R,1}(\varnothing,\varnothing)\coloneqq[E_8] 1 \underbrace{2\cdots 2}_{N_L+N_R-1 \text{ curves}} 1 [E_8]  
    \qquad \longleftrightarrow \qquad
    [N_{f}= 16]\stackon{$0$}{$\mathfrak{sp}_{N_L+N_R}$}[N_{\Lambda^2}= 1] \,,
\end{equation}
that connects the undecorated $\mathcal{K}_{N_L,N_R,1}(\varnothing,\varnothing)$ theory, constructed as the fusion of two rank $N_L$ and $N_R$ E-string theories \cite{Lawrie:2023uiu} (i.e.\ generalised $E_8$ instantons), with the T-dual $\mathrm{Spin}(32)/\mathbb{Z}_2$ theory of $N_L+N_R$ Type~I instantonic branes. The T-duality invariants of \eqref{eqn:T-duality_Invariants} for the theories in \eqref{eqn:Starting_T-duality} read:
\begin{equation}
       \mathrm{dim} \ \mathcal{C} = N_L+N_R \quad , \quad \kappa_R=N_L+N_R+1 \quad , \quad  \kappa_{\mathscr{P}}=2 \quad , \quad \mathrm{rank}(\mathfrak{f})=18 \,,
\end{equation}
where we omitted the structure constant $\kappa_{F}$ as it evaluates to $1$ in any case.

In general, we can extend the aforementioned duality in two possible ways: firstly, one straightforwardly notices that the following two theories
\begin{equation}\label{eqn:T-dual_Zero_with_Zero}
   [N_{f}= 16]\stackon{$0$}{$\mathfrak{su}_{N_L+N_R+1}$}[N_{\Lambda^2}= 2] \qquad \longleftrightarrow  \qquad
   [N_{f}= 16]\stackon{$0$}{$\mathfrak{sp}_{N_L+N_R}$}[N_{\Lambda^2}= 1] \,,
\end{equation}
are T-dual by computing the quantities in \eqref{eqn:T-duality_Invariants}.

Moreover, consider the $\mathcal{K}_{N_L,N_R,k}(\varnothing,\varnothing)$ theories
\begin{equation}
    \mathcal{K}_{N_L,N_R,k}
\coloneqq [E_8] 1 2 \stackon{$2$}{$\mathfrak{su}_2$}\cdots \stackon{$2$}{$\mathfrak{su}_k$} \underbrace{\stackon{$2$}{$\mathfrak{su}_k$} \cdots \stackon{$2$}{$\mathfrak{su}_k$}}_{N_L+N_R-1} \stackon{$2$}{$\mathfrak{su}_k$} \cdots \stackon{$2$}{$\mathfrak{su}_2$} 2 1 [E_8] \,,
\end{equation}
defined via the trivial embedding $\varnothing$ of $\mathbb{Z}_k$ inside $E_8$,  and compute the T-dual invariants:
\begin{equation}
       \mathrm{dim} \ \mathcal{C} = k(N_L+N_R+1) + k(k+1) + 1 \quad , \quad \kappa_R=\mathrm{dim} \ \mathcal{C} +1 \quad , \quad  \kappa_{\mathscr{P}}=2 \quad , \quad \mathrm{rank}(\mathfrak{f})=18 \,.
\end{equation} 
Next, tune a $0$-curve model such that it produces a candidate T-dual theory:
\begin{equation}\label{eqn:T-dual_General_k_with_Zero}
        \mathcal{K}_{N_L,N_R,k}(\varnothing,\varnothing)  
        \qquad \longleftrightarrow \qquad
        [N_{f}= 16]\stackon{$0$}{$\mathfrak{su}_{k(N_L+N_R+1) + k(k+1) + 2}$}[N_{\Lambda^2}= 2] \ .
\end{equation}
The consequences of this multiple chain of dualities among \eqref{eqn:Starting_T-duality}, \eqref{eqn:T-dual_Zero_with_Zero}, and \eqref{eqn:T-dual_General_k_with_Zero}, that we refer to as \textit{$0$-curve correspondence}.

Firstly, by the analysis of \cite{Lawrie:2023uiu} we can extract how the Higgs branch RG-flow affects the structure constant $\kappa_R$, and therefore $\mathrm{dim} \ \mathcal{C}$ because $\kappa_R=\mathrm{dim} \ \mathcal{C}+1$, for the $0$-curve models. The general change $\Delta \kappa_R = \ell_I (h^\vee_{\mathfrak{g}_I}-h^\vee_{\mathfrak{g}'_I})$ induced by the classical Higgs mechanism of $\mathfrak{g}_I$ breaking to $\mathfrak{g}_I'\subset \mathfrak{g}_I$, weighted by the null eigenvector entry corresponding to the compact curve $C_I$, simplifies since $\ell=(1)$. Therefore, all the possible Higgsing directions can be tabulated, see Table~\ref{tab:summary}, by the slices that they are induced from.
However, one could argue that slices leading to product theories are a source of ambiguity for the assignation of a unique $\Delta \kappa_R$ and are symptoms of a violation of the monotonicity theorem derived in \cite{Lawrie:2023uiu}, but in the hypothesis in which the theorem has been stated the basis cannot split in multiple theories. Moreover, the T-duality statement only regards non-product theories, thus the monotonicity argument and the usefulness of the Hasse diagram in determining T-dual models are safeguarded by restricting the subset of nodes determined by single-component theories.

This result, compared to what was found in \cite{Lawrie:2023uiu} and together with \eqref{eqn:T-dual_General_k_with_Zero}, establishes that for every choices of homomorphism $\rho_L$ and $\rho_R$ of $\mathcal{K}_{N_L,N_R,k}(\rho_L,\rho_R)$ there exists a putative T-dual $0$-curve model obtained by Higgsing the original relation \eqref{eqn:T-dual_General_k_with_Zero} and tracking $\kappa_R$ along the RG-flow. 

The second reason why the $0$-curve correspondence leads to interesting insights lies in the fact that we always have good magnetic quivers, given by the special unitary decorated $0$-curve model, T-dual to each of the $\mathcal{K}_{N_L,N_R,k}(\rho_L,\rho_R)$ theories. In fact, in \cite{DelZotto:2023nrb} it has been shown that this was not always the case when considering more involved $\mathrm{Spin}(32)/\mathbb{Z}_2$ T-duals to the $E_8 \times E_8$ framework. The remarkable property of these magnetic quivers is that they share the same shape:
\begin{equation}\label{eqn:MQ-K_{N_L,N_R,k}(general)}
\begin{gathered}
\includegraphics[width=0.9\textwidth, page=38]{pics/0-curve_figures.pdf}   
\end{gathered}
\end{equation}
this quiver \eqref{eqn:MQ-K_{N_L,N_R,k}(general)} encapsulates the magnetic quiver of a $\mathcal{K}_{N_L,N_R,k}(\rho_L,\rho_R)$ LST for every homomorphisms choice\footnote{We refer to \cite{Lawrie:2023uiu} for the dictionary between homomorphisms, instantonic branes, and node labels.}, and it is clearly in correspondence with the ones drawn in \eqref{eq:MQ_SU-even} and \eqref{eq:MQ_SU-odd}.

The same shape of all these quivers for T-dual models leads to a proposal for a method to infer the maximal flavour symmetry algebra we expect after compactification. From magnetic quivers, one can read the realised flavour symmetry via the \emph{set of balanced nodes}, and this should hold also for a putative magnetic quiver theory for the compactified theory. Hence, we propose that the maximal non-Abelian 5d flavour symmetry can be read from the common subset of balanced nodes of the magnetic quivers associated with the T-dual six-dimensional theories.
\newtheorem*{theorem*}{Algorithm}
\begin{theorem*}[\textbf{5d Maximal Non-Abelian Symmetry Rule for LSTs}]
    The maximally realised non-Abelian flavour symmetry in the 5d theory obtained by compactifying a 6d LST model or its T-duals is the common subset of balanced nodes shared from the various magnetic quivers of the theory and its duals.
\end{theorem*}

A first non-trivial result that can be extracted with this rule is exactly in the case of \eqref{eqn:Starting_T-duality}. In detail, the magnetic quiver for the $\mathcal{K}_{N_L,N_R,1}(\varnothing,\varnothing)$ theory is:
\begin{equation}\label{eqn:MQ-K_{N_L,N_R,1}(1,1)}
\begin{gathered}
\includegraphics[width=0.9\textwidth, page=39]{pics/0-curve_figures.pdf} \,.  
\end{gathered}
\end{equation}
Thus, applying the proposed algorithm to \eqref{eqn:MQ-K_{N_L,N_R,1}(1,1)} and the corresponding \eqref{eq:MQ_SU-even} or \eqref{eq:MQ_SU-odd} magnetic quiver dictated by \eqref{eqn:Starting_T-duality}, results in a 5d non-Abelian symmetry $\mathfrak{f}^{5d}_{n.a.}= \mathfrak{su}_8 \oplus \mathfrak{su}_8$, which for $N_L=N_R=1$ enhances to $\mathfrak{f}^{5d}_{n.a.}= \mathfrak{su}_{16}$. 

Another surprising application constitutes the following T-dual theories: the heterotic $E_8 \times E_8$ model with non-Abelian symmetry $\sorm(16) \times \sorm(16)$, 
\begin{equation}
    \mathcal{K}_{N_L,N_R,2p}\left( (2')^p, (2')^p \right) \coloneqq \ [N_f=8] \stackon{$1$}{$\mathfrak{sp}_p$} \underbrace{\stackon{$2$}{$\mathfrak{su}_{2p}$} \cdots \stackon{$2$}{$\mathfrak{su}_{2p}$}}_{N_L+N_R-1} \stackon{$1$}{$\mathfrak{sp}_{p}$} [N_f=8] \,, 
\end{equation}
with the T-dual invariant quantities  \eqref{eqn:T-duality_Invariants} given by
\begin{equation}
       \mathrm{dim} \ \mathcal{C} = (2p+1)(N_L+N_R) \quad , \quad \kappa_R=\mathrm{dim} \ \mathcal{C} +1 \quad , \quad  \kappa_{\mathscr{P}}=2 \quad , \quad \mathrm{rank}(\mathfrak{f})=18 \,,
\end{equation}
and the $0$-curve model
\begin{equation}\label{eqn:T-dualSO(16)xSO(16)with0}
    \mathcal{K}_{N_L,N_R,2p}\left( (2')^p, (2')^p \right) 
    \qquad \longleftrightarrow \qquad
    [N_{f}= 16]\stackon{$0$}{$\mathfrak{su}_{(2p+1)(N_L+N_R) + 1}$}[N_{\Lambda^2}= 2] \ .
\end{equation}
This T-dual pair follows from running a Higgs branch RG-flow on the natural dual pair in \eqref{eqn:T-dual_General_k_with_Zero}, and tracking the change in the duality invariant quantities. Surprisingly such a theory admits another $\mathrm{Spin}(32)/\mathbb{Z}_2$ dual \cite{DelZotto:2022ohj} with curve configuration
\begin{equation}
  \widetilde{\mathcal{K}}_{N}\big( \underbrace{8 , 0 , \cdots, 0,  8}_{p} ; \mathfrak{su}_{2p} \big) :=  \ [N_f=8] \stackon{$1$}{$\mathfrak{sp}_N$} \underbrace{\stackon{$2$}{$\mathfrak{su}_{2N}$} \cdots \stackon{$2$}{$\mathfrak{su}_{2N}$}}_{p-1} \stackon{$1$}{$\mathfrak{sp}_N$} [N_f=8] \,, 
\end{equation}
where $N=N_L+N_R$, with non-Abelian $
\sorm(16) \times \sorm(16)$ flavour symmetry.

It has been a long-standing belief that that since the algebra $\mathfrak{so}_{16}\oplus \mathfrak{so}_{16}$ is a subalgebra of both the Heterotic $E_8\times E_8$ and $\mathrm{Spin}(32)/\mathbb{Z}_2$ models, no Wilson lines need to be turned on during the compactification. Conversely from the 5d Maximal Non-Abelian Symmetry algorithm, introduced in this section and applied on \eqref{eqn:T-dualSO(16)xSO(16)with0}, it seems that some Wilson lines need to be turned on in order to obtain an $\mathfrak{f}^{5d}_{n.a.}= \mathfrak{su}_8 \oplus \mathfrak{su}_8$ as the magnetic quiver for $\mathcal{K}_{N_L,N_R,2p}\left( (2')^p, (2')^p \right)$ is:
\begin{equation}\label{eqn:MQ-K_{N_L,N_R,2p}(2p,2p)}
\begin{gathered}
\includegraphics[width=0.9\textwidth, page=18]{pics/0-curve_figures.pdf} \,.  
\end{gathered}
\end{equation}
This fact is in agreement with \cite{McInnes:1999va}: by analysing the global structure of the $\mathfrak{so}_{16}\oplus \mathfrak{so}_{16}$ subalgebra of the Heterotic models, it was realised that the groups it lifts to are different depending on which frame one considers. Consequently, as a subgroup common to both the $E_8 \times E_8$ and the $\mathrm{Spin}(32)/\mathbb{Z}_2$ Heterotic strings does not exist, Wilson lines need to be turned on --- possibly breaking the 6d algebra to what was found via the algorithm.

We remark that although this algorithm seems a promising way to peek at the features of the compactified models only an explicit computation of the 5d theory can shed light on the T-duality statement in all its aspects. 

\section{Conclusions}
\label{sec:conclusions}
The Higgs branches of six-dimensional $\mathcal{N}=(1,0)$ little string theories provide valuable insights for both physics and mathematics. The Higgs branch analysis offers new information on T-dual candidates and the magnetic quiver realisation allows for the extraction of potential new symplectic singularities.

As we have shown in this paper, the little string models realised on a zero curve equipped with a special unitary or symplectic gauge group can be studied via Type IIA compactification of M5 branes probing a singular $\mathbb{C}^2/\mathbb{Z}_k$ space within two M9 walls.
A magnetic quiver can be read off for each of these theories. For the $\surm$-type LSTs, the magnetic quiver is ``good'': \eqref{eq:MQ_SU-even} for $\surm(N)$ with $N=2\ell$ even,  and \eqref{eq:MQ_SU-odd} for $\surm(N)$ with $N=2\ell+1$ odd.
In contrast, for the $\sprm(\ell)$ theories on a single $0$-curve, the magnetic quiver \eqref{eq:MQ_Sp} is ``bad''.
For the special SU/Sp LSTs, all the magnetic quivers \eqref{eq:MQ_SU6_18fund}, \eqref{eq:MQ_SU6_17fund}, \eqref{eq:MQ_Sp3_17-half_fund} are ``good''.

\begin{table}[ht]
    \centering
    \ra{2}
    \begin{tabular}{lcc}
        \toprule 
        Higgs branch RG-flow &   transition & $\Delta \kappa_R$ \\ \midrule
        $\scriptstyle{[N_f{=}16]}\overset{\surmL(N)}{0} \scriptstyle{[N_{\Lambda^2}{=}2]} 
        \; \longrightarrow \;
        \scriptstyle{[N_f{=}16]}\overset{\surmL(N-1)}{0} \scriptstyle{[N_{\Lambda^2}{=}2]} $ & $\begin{cases} a_{15} \,, & N\geq 4 \\ a_{17} \,, & N=3   \end{cases} $ & $-1$  \\
        $\scriptstyle{[N_f{=}16]}\overset{\surmL(2\ell)}{0} \scriptstyle{[N_{\Lambda^2}{=}2]} 
        \; \longrightarrow \; 
        \scriptstyle{[N_f{=}16]}\overset{\sprmL(\ell)}{0} \scriptstyle{[N_{\Lambda^2}{=}1]} $
        &  $h_{2,\ell}$ & $-\ell+1$ \\
        $\scriptstyle{[N_f{=}16]}\overset{\surmL(N)}{0} \scriptstyle{[N_{\Lambda^2}{=}2]} 
        \; \longrightarrow \;
        \scriptstyle{[N_f{=}16]}\overset{\sprmL(k)}{0} \scriptstyle{[N_{\Lambda^2}{=}1]}
         \quad \sqcup \quad
        \scriptstyle{[N_f{=}16]}\overset{\surmL(N-2k)}{0} \scriptstyle{[N_{\Lambda^2}{=}2]} $
        &  $A_{1}$ & $-k+1$ \\\midrule
         
        $\scriptstyle{[N_f{=}16]}\overset{\sprmL(1)}{0} \scriptstyle{[N_{\Lambda^2}{=}1]} 
        \; \longrightarrow \;
       \overset{\emptyset}{0}$
        &  $d_{16}$ & $-2$ \\
        
        $\scriptstyle{[N_f{=}16]}\overset{\sprmL(\ell)}{0} \scriptstyle{[N_{\Lambda^2}{=}1]} 
        \; \longrightarrow \;
        \scriptstyle{[N_f{=}16]}\overset{\sprmL(k)}{0} \scriptstyle{[N_{\Lambda^2}{=}1]}
        \quad \sqcup \quad
        \scriptstyle{[N_f{=}16]}\overset{\sprmL(\ell-k)}{0} \scriptstyle{[N_{\Lambda^2}{=}1]} $
        &  $\begin{cases} A_1 \, , & \ell-k=k \\ 
        m \, , & \ell -k \neq k\end{cases}$ & $+1$ \\ \midrule
        %
         $\scriptstyle{[N_f{=}18]}\overset{\surmL(6)}{0} \scriptstyle{[N_{\Lambda^3}{=}1]} 
        \; \longrightarrow \;
        \scriptstyle{[N_f{=}16]}\overset{\surmL(5)}{0} \scriptstyle{[N_{\Lambda^2}{=}2]} $ & $a_{17}$ & $-1$  \\ 
         $\scriptstyle{[N_f{=}18]}\overset{\surmL(6)}{0} \scriptstyle{[N_{\Lambda^3}{=}1]} 
        \; \longrightarrow \;
        \scriptstyle{[N_f{=}16]}\overset{\sprmL(3)}{0} \scriptstyle{[N_{\Lambda^2}{=}1]} 
         \quad \sqcup \quad
        \scriptstyle{[N_f{=}16]}\overset{\sprmL(3)}{0} \scriptstyle{[N_{\Lambda^2}{=}1]} $ & $A_{1} $  & $+2$ \\ \midrule
         $\scriptstyle{[N_f{=}17]}
         \underset{[N_{\Lambda^2}{=}1]}{\overset{\surmL(6)}{0}}
         \scriptstyle{[N_{\Lambda^3}{=}\frac{1}{2}]} 
        \; \longrightarrow \;
        \scriptstyle{[N_f{=}16]}\overset{\surmL(5)}{0} \scriptstyle{[N_{\Lambda^2}{=}2]} $ & $a_{16} $ & $-1$  \\ 
         $\scriptstyle{[N_f{=}17]} \underset{[N_{\Lambda^2}{=}1]}{\overset{\surmL(6)}{0}}\scriptstyle{[N_{\Lambda^3}{=}\frac{1}{2}]} 
        \; \longrightarrow \;
        \scriptstyle{[N_f{=}16]}\overset{\sprmL(3)}{0} \scriptstyle{[N_{\Lambda^2}{=}1]} 
        \quad \sqcup \quad
        \scriptstyle{[N_f{=}17+\tfrac{1}{2}]}\overset{\sprmL(3)}{0} \scriptstyle{[N_{\Lambda^3}{=}\frac{1}{2}]} $ & $A_{2}$ & $+2$  \\  \midrule
         $\scriptstyle{[N_f{=}17+\frac{1}{2}]}\overset{\sprmL(3)}{0} \scriptstyle{[N_{\Lambda^3}{=}\frac{1}{2}]} 
        \; \longrightarrow \;
        \scriptstyle{[N_f{=}16]}\overset{\sprmL(2)}{0} \scriptstyle{[N_{\Lambda^2}{=}1]} $ & $b_{17}$  & $-1$ \\ \bottomrule
    \end{tabular}
    \caption{Summary of the Higgs branch RG-flows of the LSTs defined on a single curve of self-intersection $0$ with fiber of type $I$.}
    \label{tab:summary}
\end{table}

Using branching rules, brane dynamics, quiver subtraction, the decay and fission algorithm, and F-theory arguments, we identified a plethora of Higgs branch RG-flows; these are summarised in Table~\ref{tab:summary}. The agreement between the different techniques constitutes a non-trivial consistency check of the results.

Building on the magnetic quiver tools --- such as quiver subtraction and the decay and fission algorithm ---  one can derive all the slices that can appear in the Higgs branch Hasse diagram; thus, inferring the change in the structure constant characterising the $2\text{-}\mathrm{Group}$ structure properties of LSTs. Therefore, matching the T-dual invariant quantities \eqref{eqn:T-duality_Invariants} between different little string models allows us to infer candidate T-duals, and leads to the claim that every linear LST comprised of $-1$ and $-2$ curves equipping only special unitary or symplectic algebras is T-dual to a zero curve model equipped with one of the aforementioned algebras.

Moreover, these results lead us to conjecture an algorithm to predict the flavour symmetry of the five-dimensional theory that emerges from the compactification with Wilson lines of the six-dimensional T-dual models. The key point is that the shape of all the good magnetic quivers for the T-dual theories is the same; hence, one can predict the non-Abelian part of the five-dimensional flavour symmetry by inspecting the common subset of balanced nodes in all the models. Significantly, this algorithm correctly predicts the presence of Wilson lines in Heterotic models with $\mathfrak{so}_{16}\oplus \mathfrak{so}_{16}$ flavour symmetry, as dictated by the different global group structure.

\paragraph{Outlook.} While the six-dimensional little string theories have been explored via multiple approaches, it still remains an open problem to write down the five-dimensional compactification of these theories and the specific set of Wilson lines needed on each T-dual model.

The inferences made so far about dual candidates and flavour symmetry may be demystified by such approach or may work as an hint to shed light on what to expect from such a computation.

Furthermore, it is important to extend the setup by inclusion of $\mathrm{O6}$ planes in the direction of D6 planes. The magnetic quivers for such theories are expected to follow arguments along the lines of \cite{Cabrera:2019dob,Sperling:2021fcf,Hanany:2022itc}. However, the analysis of the Higgs branch singularity structure and the RG-flows it encodes requires further research into the magnetic quiver techniques.

\paragraph{Acknowledgements.}
We are grateful to Florent Baume, Antoine Bourget, Julius Grimminger, Amihay Hanany, Craig Lawrie, and Zhenghao Zhong for insightful discussions.
LM thanks the Mathematical Physics Group, University of Vienna for warm hospitality and support. LM acknowledges support from DESY (Hamburg, Germany), a member of the Helmholtz Association HGF.
The work of MS is supported by Austrian Science Fund (FWF), START project STA 73-N. MS also acknowledges support from the Faculty of Physics, University of Vienna.
We thank the \emph{Workshop on Symplectic Singularities and Supersymmetric QFT} (July 2023, Amiens) for partial support and a stimulating environment.

 \appendix

\section{Structure of monopole operators}
\label{app:monopole}
While studying zero curve models, F-theory can also help predicting (a subset of) the expected flavour symmetry for the theory. In fact, considering the following LST:
\begin{equation}
     [N_{f}= 16]\stackon{$0$}{$\mathfrak{su}_N$}[N_{\Lambda^2}= 2] \,,
\end{equation}
for $N \ge 5$, all the representation are complex, thus we expect an
\begin{equation}\label{eqn:Flavour-From-F-theory}
    \mathfrak{f}=\mathfrak{s}\left(\mathfrak{u}_{16} \oplus \mathfrak{u}_2 \right)\cong \mathfrak{su}_{16} \oplus \mathfrak{su}_2 \oplus \mathfrak{u}_1 
\end{equation}
flavour symmetry rotating the hypermultiplets. The surviving $\mathfrak{u}_1$ can be tracked down by cancellation of the ABJ anomaly with the counting technique introduced in \cite{Apruzzi:2020eqi} as the Green–Schwarz mechanism \cite{Green:1984sg,Sagnotti:1992qw} does not apply to zero curve models, thus allowing the formulation of an anomaly polynomial for this particular little string theory \cite{Cordova:2020tij}. The generator $T_{\mathrm{survivor}}$ of the surviving $\mathfrak{u}_{1}$ factor is explicitly:
\begin{equation}
    T_{\mathrm{survivor}} \propto \left(N-4\right)t_{\mathrm{fund}} - 16  \, t_{\Lambda^2} \,,
\end{equation}
where we labelled $t_{\cdot}$ the generator for each of the $\mathfrak{u}_1$ factors acting singularly respectively on the fundamental or the $2^{\mathrm{nd}}$-rank anti-symmetric hypermultiplets.\\
Since we also have a good magnetic quiver for this model, one could check whether \eqref{eqn:Flavour-From-F-theory} reproduce the full flavour symmetry of the theory or if the geometry is hiding some Abelian factors. From Hilbert Series computation on \eqref{eq:MQ_SU-even} and \eqref{eq:MQ_SU-odd} we obtain:
\begin{equation}
    \mathrm{HS}\left( t \right)= 1 + 260 \, t^2 + \cdots \ .
\end{equation}
This result clearly stems surprise as one would expect a contribution of $255$ coming from the $ \mathfrak{su}_{16}$, as also expected from the presence of the $a_{15}$ slice in the Hasse diagram, plus a factor of $4$ obtained from the $ \mathfrak{su}_{2}\oplus \mathfrak{u}_{1}$ factors also seen as the symmetry of the $h_{2,N}$ slice. The missing contribution comes from an extra $\mathfrak{u}_{1}$ factor that stems from particular choice of fluxes that we are about to elucidate. The physical explanation of this extra symmetry factors can be tracked from the brane construction as the preserved symmetry of the half NS5s inside the O8 orbifolds, as in the M-theory uplift of \cite{DelZotto:2023nrb}.

From \cite{DianMansi} emerged that the structure of the magnetic fluxes contributing at a certain R-charge level is made of \textit{connected components} and \textit{disconnected} ones. Let us give here a brief insight of what it is meant by that. Recall that a Coulomb branch Hilbert series for a $3d$ $\mathcal{N}=4$ quiver theory \cite{Cremonesi:2013lqa} is of the form:
\begin{equation}
    \mathrm{HS}(t)=\sum \limits_{m \in \Lambda^{*}_{\hat{G}}/ \mathcal{W}_{\hat{G}}} P_G\left(t,m\right) t^{2\Delta(m)} \ ,
\end{equation}
where $P_G\left(t,m\right) $ is the product of the polynomials counting the Casimir invariants of each residual gauge group for the choice of fluxes $m$. The $\Delta(m)$ is the R-charge of the monopole operator of GNO charge $m$, thus $2\Delta(m)\in \mathbb{N}$, and is given by: 
\begin{equation}\label{eqn:R-chargeMonopole}
    \Delta(m)=   - \sum \limits_{\alpha \in \Delta_{+}} | \alpha(m) | + \frac{1}{2} \sum \limits_{j=1}^n \sum \limits_{\rho_{i}\in \mathcal{R}_i} | \rho_{i}\left( m \right) |  \ ,
\end{equation}
with $\Delta_{+}$ being the set of positive root of the gauge algebra and $\rho_i$ is the weight of the hypermultiplet $i$ in the $\mathcal{R}_i$ representation. 
As such the conformal dimension in \eqref{eqn:R-chargeMonopole} as well as the classical factor $P_G\left(t,m\right) $ are well under control for each choice of fluxes and allow for a systematic understanding in terms of \textit{simple moves} of how a certain configuration contributes at a fixed R-charge level. In fact, for a good quiver theory with $N$ gauge nodes one can argue as follows:  
\begin{itemize}
    \item Start by considering the $0$ level, here the only contribution is given by a trivial choice of fluxes for every gauge node in the quiver.
    \item Now moving to level $2$ is achieved performing a simple move, i.e. turning on the flux $m_i=\{1,0,\cdots,0\}$ on the $i$-th balanced gauge node.
    \item Next, one can remain on the same level turning on a flux $m_j=\{1,0,\cdots,0\}$ in the adjacent balanced node. And this concludes the analysis on the connected component in the restricted space $M_\ge=\{ m_i  \text{ with } i \in \{1,\cdots, N\} \ \mathrm{s.t.} \ \left(m_i\right)_j \ge 0 \ \forall j \}$.
    \item All the other choice of fluxes in $M_\ge$ contributing to the adjoint action that cannot be obtained with these moves are called disconnected.
\end{itemize}
Of course each configuration in $M_\ge$ contributes to $2$ distinct configuration as the conformal dimension in \eqref{eqn:R-chargeMonopole} is invariant under flipping of all fluxes' signs. Plus one needs to remember that together with the monopole contributions there is also a $t^2$ term in the expansion of the Casimir invariants that accounts for an addend equal to the number of gauge nodes in the quiver after ungauging.

Given a quiver shaped as a doubly overextended affine D-type Dynkin, as considered in this paper, with gauge nodes 

\begin{equation}
\begin{gathered}
   \includegraphics[scale=1.5, page=33]{pics/0-curve_figures.pdf} \ ,
\end{gathered} 
\end{equation}
we can label each node as
\begin{equation}
\begin{gathered}
   \includegraphics[scale=1.5, page=34]{pics/0-curve_figures.pdf} \ ,
\end{gathered} 
\end{equation}
and apply the aforementioned rules. The connected monopole configurations one can extract at level $2$ are
\begin{itemize}
    \item $k=\mathrm{even}$, \begin{enumerate}
    \item $m_i=(1,0,\cdots, 0)$ with $i\in \{1,\cdots,N-1\}$
    \item $m_i=(1,\cdots, 0)$ and $m_{i+1}=(1,\cdots, 0)$ with $i\in \{1,\cdots,N-2\}$, for a total of $N-2$ configurations.
    \item $m_i=(1,\cdots, 0)$ and all $m_{i+j}=(1,\cdots, 0)$ with $j\in \{1,\cdots,k\}$ $i\in \{1,\cdots,N-1-k\}$, for a total of $N-1-k$ configurations.
\end{enumerate}
    \item $k=\mathrm{odd}$, \begin{enumerate}
    \item $m_i=(1,0,\cdots, 0)$ with $i\in \{2,\cdots,N-2,N,N+1\}$
    \item $m_i=(1,\cdots, 0)$ and $m_{i+1}=(1,\cdots, 0)$ with $i\in \{2,\cdots,N-3\}$, plus the two configurations $m_{N}=m_{2}=(1,\cdots, 0)$ and $m_{N-2}=m_{N+1}=(1,\cdots, 0)$, for a total of $N-1$ configurations.
    \item $m_i=(1,\cdots, 0)$ and all $m_{i+j}=(1,\cdots, 0)$ with $j\in \{1,\cdots,k\}$ $i\in \{2,\cdots,N-2-k\}$,plus the two configurations of length $k$ stemming from the $N$-th and the $N+1$-th nodes, for a total of $N-1-k$ configurations. Notice that for $k=N-1$ these two extra configuration coalesce and count as 1.
\end{enumerate}
\end{itemize}
In both cases one is correctly led, upon remembering that each configuration counts twice, to the grand total of $N(N-1)$, to which one should add the number of $t^2$ coefficients coming from the Casimir invariants, i.e.\ the number of gauge nodes after ungauging:
\begin{equation}
    N(N-1)+N+3-1=N^2+2=\left(N^2-1 \right) + 3 \,.
\end{equation}
This result reproduces the one expected for an electric theory alike to the $0$-curve one having an $\mathfrak{su}_N$ flavour symmetry rotating its fundamental hypermultiplets and an $\mathfrak{su}_2$ one related to the anti-symmetric matter.

The extra disconnected monopole configurations have equal structure for every $k$ and correspond to the following monopole flux structure:
\begin{equation}
    m_1=m_{N-1}=m_N=m_{N+1}=(1,0,\cdots,0) \text{ and } m_i=(1,1,0,\cdots,0) \ \forall i \neq \{1,N-1,N,N+1\} \ ,
\end{equation}
and they contribute as two extra $\mathfrak{u}_1$ factors that correctly reproduce the Hilbert series
\begin{equation}
    \mathrm{HS}(t,N)=1+ \left( N^2 +4 \right) t^2 + \cdots \ ,
\end{equation}
identified for the zero curve theory with $N=16$.

\FloatBarrier

\bibliographystyle{JHEP}
\bibliography{bibliography.bib}
\end{document}